\documentclass[aps,prx,twocolumn,amsmath,amssymb,showpacs,superscriptaddress,longbibliography]{revtex4-1}

\usepackage{graphicx}
\usepackage{dcolumn}
\usepackage{bm}
\usepackage{color}

\begin{document}

\title{Self-organized explosive synchronization in complex networks: \\Emergence of synchronization bombs}

\author{Llu\'is Arola-Fern\'andez}
\affiliation{Departament d'Enginyeria Inform\`{a}tica i Matem\`{a}tiques, Universitat Rovira i Virgili, 43007 Tarragona, Spain}%

\author{Sergio Faci-L\'azaro}
\affiliation{Department of Condensed Matter Physics, University of Zaragoza, 50009 Zaragoza (Spain).}
\affiliation{GOTHAM lab, Institute of Biocomputation and Physics of Complex Systems, University of Zaragoza, 50018 Zaragoza, Spain}%
 
 \author{Per Sebastian Skardal}
\affiliation{Department of Mathematics, Trinity College, Hartford, CT 06106, USA}%

\author{Emanuel-Cristian Boghiu} 
\affiliation{ICFO –- Institut de Ci\`{e}ncies Fot\`{o}niques, The Barcelona Institute of Science and Technology, 08860 Castelldefels (Barcelona), Spain} 

\author{Jes\'us G\'omez-Garde\~nes}%
\thanks{gardenes@unizar.es}
\affiliation{Department of Condensed Matter Physics, University of Zaragoza, 50009 Zaragoza (Spain).}
\affiliation{GOTHAM lab, Institute of Biocomputation and Physics of Complex Systems, University of Zaragoza, 50018 Zaragoza, Spain}%

\author{Alex Arenas}
\email{alexandre.arenas@urv.cat}
\affiliation{Departament d'Enginyeria Inform\`{a}tica i Matem\`{a}tiques, Universitat Rovira i Virgili, 43007 Tarragona, Spain}%

\date{\today}

\begin{abstract}

\emph{We introduce the concept of synchronization bombs as large networks of coupled heterogeneous oscillators that operate in a bistable regime and abruptly transit from incoherence to phase-locking (or vice-versa) by adding (or removing) one or a few links. Here we build a self-organized and stochastic version of these bombs, by optimizing global synchrony with decentralized information in a competitive link-percolation process driven by a local rule. We find explosive fingerprints on the emerging network structure, including frequency-degree correlations, disassortative patterns and a delayed percolation threshold. We show that these bomb-like transitions can be designed both in systems of Kuramoto --periodic-- and R\"ossler --chaotic-- oscillators and in a model of cardiac pacemaker cells. We analytically characterize the transitions in the Kuramoto case by combining a precise collective coordinates approach and the  Ott-Antonsen ansatz. Furthermore, we study the robustness of the phenomena under changes in the main parameters and the unexpected effect of optimal noise in our model. Our results propose a minimal self-organized mechanism of network growth to understand and control explosive synchronization in adaptive biological systems like the brain and engineered ones like power-grids or electronic circuits. From a theoretical standpoint, the emergence of synchronization explosions and bistability induced by localized structural perturbations --without any fine-tuning of global parameters-- joins explosive synchronization and percolation under the same mechanistic framework.}

\end{abstract}

\maketitle


\section{\label{sec:introduction}Introduction}

The emergence of abrupt, explosive transitions in the macroscopic behavior of complex networked systems is a fascinating phenomenon, ubiquotuos in fields ranging from neuroscience to biology and engineering. There is increasing empirical evidence that explosive synchronization in brain activity is associated with the onset of anesthesic-induced unconsciousness \cite{joiner13,kim16,kim17}, epileptic seizures \cite{wang17a,wang17b} and fibromyalgia \cite{lee18} and it explains biological switches displaying abrupt responses to external perturbations \cite{chatterjee08}. Also, in infrastructural and power-grids networks, it is crucial to detect and control small vulnerabilites that can lead to abrupt structural damages and global desynchronization blackouts \cite{dobson07,newman10}. 

From a theoretical perspective, explosive percolation --an abrupt growth of the giant component of the network induced by the addition or removal of single links-- was found to occur when competitive rules are applied on the choice of the links in a way that the formation of a giant cluster is delayed \cite{achlioptas09}. The discovery of this abrupt structural transition, which was shown to be continuous in the thermodynamic limit but with anomalous scaling properties, triggered further analyses to understand the mechanisms that can lead to the explosive behavior in the network growth. In parallel, abrupt transitions were explored in the collective dynamics of the system when considering a physical process among the units, as the spreading of a disease \cite{dedomenico16,bottcher15,matamalas20}, opinion diffusion \cite{gardenes16} or traffic flow \cite{echenique2005,lampo21}, to name a few \cite{dsouza19,boccaletti16}. 

A particularly suitable framework to model the birth of explosive transitions is the synchronization process of coupled oscillators. The phenomena of collective synchronization is widely spread in natural, social and technological systems \cite{pikovsky01,arenas08}. Its ubiquity has attracted the interest of the physics community, that have tackled its study through minimal models that capture the transition between a disordered phase and coherent dynamics. For populations of heterogeneous phase-oscillators coupled all-to-all \cite{arenas08,pikovsky01}, abrupt transitions in synchrony as the coupling parameter is increased were found to occur for a uniform distribution of frequencies \cite{pazo05} and the hysteresis cycle involving incoherence and partial synchrony was exactly characterized for a bimodal distribution with a shallow dip \cite{martens09}. However, the bistable nature of explosive synchronization -an abrupt jump from incoherence to global synchrony induced by a change in the coupling parameter among the units, with an associated hysteresis cycle- was firstly discovered for scale-free networks (i.e. networks with very heterogeneous degree distributions) in the presence of positive correlations between the internal frequencies and the nodal degrees \cite{gardenes11}. Further analyses showed that this is only one of the possible mechanisms that inhibit the emergence of a large synchronization cluster and it was found that, by imposing frequency anti-correlations among connected units in the form of frequency gaps \cite{leyva13} or adaptive anti-Hebbian rules for the weights \cite{avalos18}, explosive transitions occur as the coupling constant is tuned. Recently, it has been found that these degree and frequency correlations associated to explosive behavior also optimize the global phase synchronization in the system \cite{arola21,wei21}. Furthermore, explosive transitions can also appear in multilayer and dynamically coupled systems \cite{zhang15,soriano19} and they can be enhanced by the presence of noise \cite{skardal14b} and higher-order -beyond pair-wise- interactions \cite{skardal19}. 

Our fundamental understanding of explosive synchronization has significantly increased in the last years, but due to the analytical challenges of synchronization dynamics on arbitrary complex networks, a rigorous framework analogous to explosive percolation is still missing \cite{dsouza19}. Importantly, explosive synchronization and percolation focus on different aspects of the system, namely in the abrupt changes on the macroscopic dynamical and structural properties, respectively, when subject to small variations of the control parameter (the coupling strength or the density of links). Interestingly, in \cite{zhang14}, the authors found that a particular choice of frequency-dependent coupling (which again induces anti-correlations) produces an explosive synchronization process where the formation of synchronized clusters is delayed analogously to its percolation counterpart.  While these results unveil a deep connection between both phenomena, the choice on the coupling dependence is heuristic and the system produces the explosive behavior under changes in a global control parameter (the coupling strength), unlike the explosive percolation which is induced locally, by adding or removing single links. 

Apart from the theoretical interest, there is an ongoing consensus that explosive synchronization phenomena is behind the operation of biological switches and neural systems \cite{chatterjee08, kim16, kim17, wang17a, wang17b, lee18, joiner13, myung18}. Biological units usually operate with limited, decentralized information and are affected by noise \cite{ishida97, izhikevich03, orlandi13, zhang16}. This ubiquity of the explosive transitions cannot be explained by means of global and deterministic optimization routes, specific network and oscillator designs or global fine-tuning of coupling parameters. Thus, it is still poorly understood how complex biological systems like the brain can self-organize to display the observed explosive behavior \cite{chatterjee08, wang17b, scarpetta18}. 

To tackle the aforementioned challenges from a theoretical perspective, here we present a self-organized dynamical network that act as a synchronization bomb, i.e. showing an abrupt synchronization transition in the course of a self-organized wiring process. This way, our model  attempts to bridge the conceptual gap between explosive synchronization and percolation by imposing local structural perturbations instead of global ones and proposes a self-organized and stochastic route to explosive synchronization by invoking a simple principle of synchrony maximization in a decentralized and noisy environment. 

The remainder of this paper is organized as follows. We first present our model of the synchronization bomb for an ensemble of Kuramoto oscillators. We introduce the optimal local rule for connecting or disconnecting units, derived from the truncated expansion of the linearized dynamics \cite{arola21} under the assumptions of maximizing global synchrony with local information, and explore the basic mechanisms and phenomenology of the synchrony-driven percolation process. Second, we analyze the explosive fingerprints that emerge on the underlying structure, in the form of degree-frequency correlations, dissasortative dynamical and structural patterns, and a delayed percolation threshold. Third, we provide an analytical characterization of the dynamics by means of the Collective Coordinates \cite{gottwald15,hancock18} (CC) and Ott-Antonsen \cite{ott08} (OA) model reduction techniques, unveiling the dependence of the main parameters and observing an excellent agreement with numerical simulations. Next, we extend the model to numerically produce synchronization bombs of coupled chaotic R\"ossler systems and cardiac pacemaker cells. We conclude with a discussion of our results and a methods' section, including the mathematical machinery used to derive the local rule and the analytical predictions for the percolation and synchronization critical thresholds. In the supplementary information (SI), we study the robustness of the presented phenomenology under variations in system parameters, and we explore in depth the effect of noisy sampling in our model, showing that the presence of noise is beneficial because it improves the decentralized optimization of synchrony driven by a local rule. 

\section{\label{sec:results}Results}

\begin{figure*}[!t]
\includegraphics[scale=0.60]{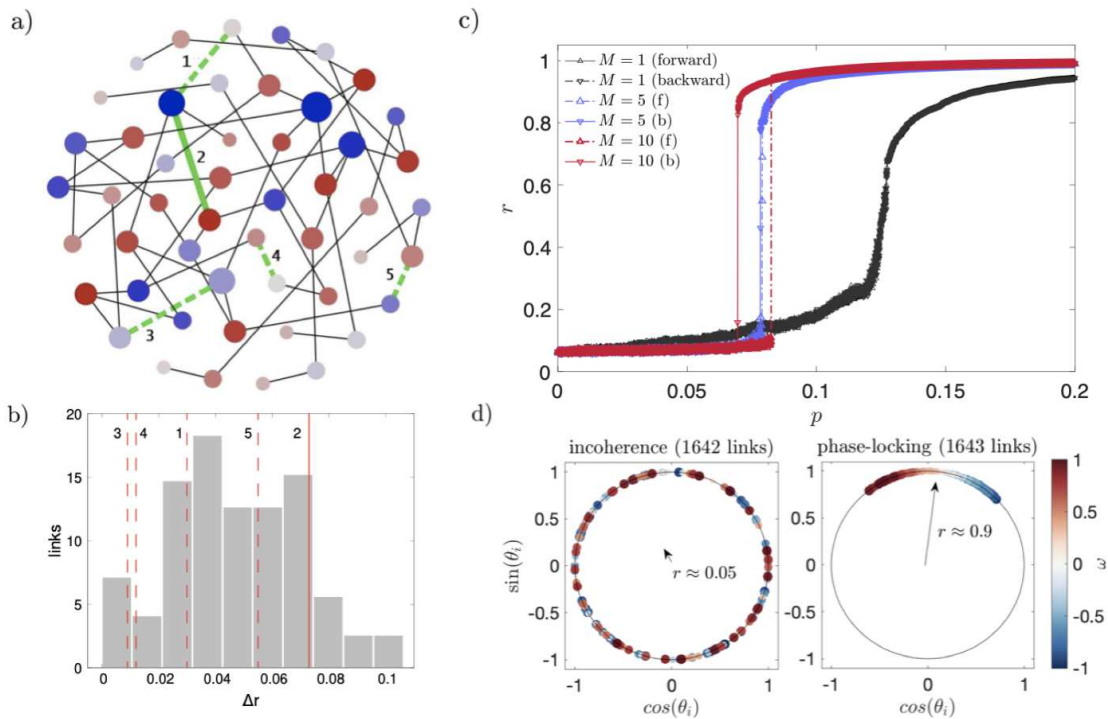}
\caption{a) Illustrative network of $N = 50$ oscillators where the size of the node is proportional to its degree, the color is related to its natural frequency (blue for $\omega=-1$, gray for $\omega=0$ and red for $\omega=1$) and the black lines represent the links between the oscillators. Green lines mark the $M=5$ potential links sampled in that $p$-step. The continuous line represents the chosen link and the dashed ones are the discarded ones. b) Histogram of the $\Delta r$ values for the existing links of the network, where red lines correspond to the values of the five sampled links. c) Example of the typical synchronization transition in our bomb-like model, with the order parameter $r$ depending on the fraction of links, $p$ in a system of $N=200$ Kuramoto oscillators for fixed coupling strength $\lambda = 0.05$ and three values of sampling $M$. Eq.(\ref{eq:kuramoto}) is numerically integrated using Heun's method, with $dt = 0.05$ and $10^4$ time steps and temporal averages of $r$ are taken at each link change. In d), we represent the oscillators phases for the $M=10$ case before (left) and after (right) the forward transition. It is important to note that this jump in the Kuramoto order parameter from incoherence to complete phase--locking takes place just with the addition of one link.}
\label{fig:illustrative} 
\end{figure*}

\textbf{\label{subsec:model}Model.} We consider a large system of heterogeneous coupled oscillators on top of a network of interactions that evolves under a competitive link percolation process \cite{achlioptas09,dsouza19}. For the dynamics, we begin with the classical Kuramoto model, a paradigmatic example of the emergence of collective synchronization \cite{kuramoto03,pikovsky01,arenas08}. An ensemble of $N$ heterogeneous Kuramoto oscillators interacting on top of a network follows the equations of motion

\begin{equation}
\dot{\theta}_i = \omega_i + \lambda\sum_{j=1}^N a_{ij} \sin(\theta_j - \theta_i), \mbox{ } \forall \ i \in 1 \dots N,
\label{eq:kuramoto}
\end{equation}

where $\theta_i$ is the phase and $\omega_i$ is the intrinsic frequency of the $i$-oscillator, $a_{ij}$ are the entries of the adjacency matrix $A$, that capture the interactions among the units and $\lambda$ is a constant coupling strength. As usual, the macroscopic behavior of the system is captured by the modulus of the Kuramoto order parameter 
\begin{equation}
r(t) = \frac{1}{N} \left|\sum_{j = 1}^N e^{i\theta_j(t)}\right|,
\label{eq:r}
\end{equation}
which measures the degree of phase synchronization and is bounded between zero and one. In the following, we will make use of temporal averages of the order parameter, i.e. $r = \langle r(t)\rangle$. Although our results can be extended to more general settings, in the following we restrict our study to the case of unweighted ($a_{ij} = 0,1$) and undirected networks $(a_{ij} = a_{ji})$, and consider, for analytical convenience, a uniform frequency distribution  $g(\omega) \in  [-\gamma,\gamma]$ with zero mean. 

The growth of the synchronization bomb is made by keeping the coupling strength $\lambda$ is constant and varying the density in the number of connections between the units, $p$, that acts as the control parameter and ranges from $0$ (disconnected network) to $1$ (fully-connected network). In the forward process we initialize our system from scratch, with a completely disconnected network $(p = 0)$ of oscillators with assigned random phases drawn from $(-\pi,\pi)$. We then run the percolation processes in which at each step one new link is added. This way the control parameter $p$ changes sufficiently slow such that the system in Eq.~(\ref{eq:kuramoto}) reaches the stationary state at each network step in the process. The addition of a new link at each $p$-step is made as follows: We uniformly sample $M$ pairs of disconnected oscillators and select the connection $(i,j)$ that maximizes the gain of synchrony given by
\begin{equation}
\Delta r_{ij} = \frac{1}{\lambda^2 N} \left( \frac{\omega_i}{k_i} - \frac{\omega_j}{k_j}\right)\left(\frac{\omega_i}{k_i^2} - \frac{\omega_j}{k_j^2}\right). 
\label{eq:rule1-cont} 
\end{equation}
In practice, when connecting isolated nodes at the very initial steps of the process, we add an infinitesimally small value to the degrees of the nodes with $k_i = 0$ to evaluate Eq.~(\ref{eq:rule1-cont}) in terms only of the natural frequencies. In the backward process, we just remove the links in the reversed order of the forward process. The proposed model is stochastic in nature but becomes completely deterministic in the limit $M \rightarrow \infty$, and it reduces to the random percolation case in $M = 1$.

Eq.~(\ref{eq:rule1-cont}) captures the actual change in the order parameter $r$ in the strong phase-locking regime (i.e. after the transition) but it can be used to estimate the impact of each link in the whole synchronization process (see Methods section \ref{subsec:rule} for in depth derivation and discussion of this expression). Note that Eq.~(\ref{eq:rule1-cont}) only exploits local information of the considered nodes, and it is maximum when the ratios frequency-degree of the nodes are large and also when their difference is large as well, pinpointing a clear signature of frequency-degree correlations and frequency anti-correlations. These correlations were imposed \emph{ad hoc} in previous models that induce explosive synchronization \cite{dsouza19}, and they could indeed emerge from applying a broader class of local percolation rules in the form $p(\omega_i,k_i,\omega_j,k_j)$, but we focus on Eq.~(\ref{eq:rule1-cont}) since it is the rule that is derived from a decentralized optimization of the phase-locking state, without other assumptions or guesses required.

In the left panels of Fig.~\ref{fig:illustrative}, we illustrate the former basic mechanics for assigning a link out of $M=5$ possible candidates. We take the forward process (construction of the network by adding links) as an example. As shown in Fig.~\ref{fig:illustrative}.a) the functional form of the basic rule, Eq.~(\ref{eq:rule1-cont}), induces some relevant features on the interplay between structural and dynamical patterns during the network growth. We observe that nodes with large (small) absolute frequencies accumulate more (less) neighbours, whereas links tend to be more present between nodes with alternate frequencies, producing bipartite-like structures, as we will explore in the following lines. In the panel Fig.~\ref{fig:illustrative}.c) we show the forward and backward explosive synchronization transitions by plotting the curves $r(p)$ when different values of $M$ are used. We observe that as $M$ increases so it does the abruptness of the transition as well as the hysteresis region. To illustrate better the explosive nature of these transitions we show in Fig.~\ref{fig:illustrative}.d) the transition from incoherence $(r \approx 0.05)$ to full phase-locking ($r \approx 0.9)$ when a unique link is added to the system. This phenomenon motivates our choice for referring to these growing networks as synchronization bombs. \\

\begin{figure*}[t!]
\includegraphics[scale=0.35]{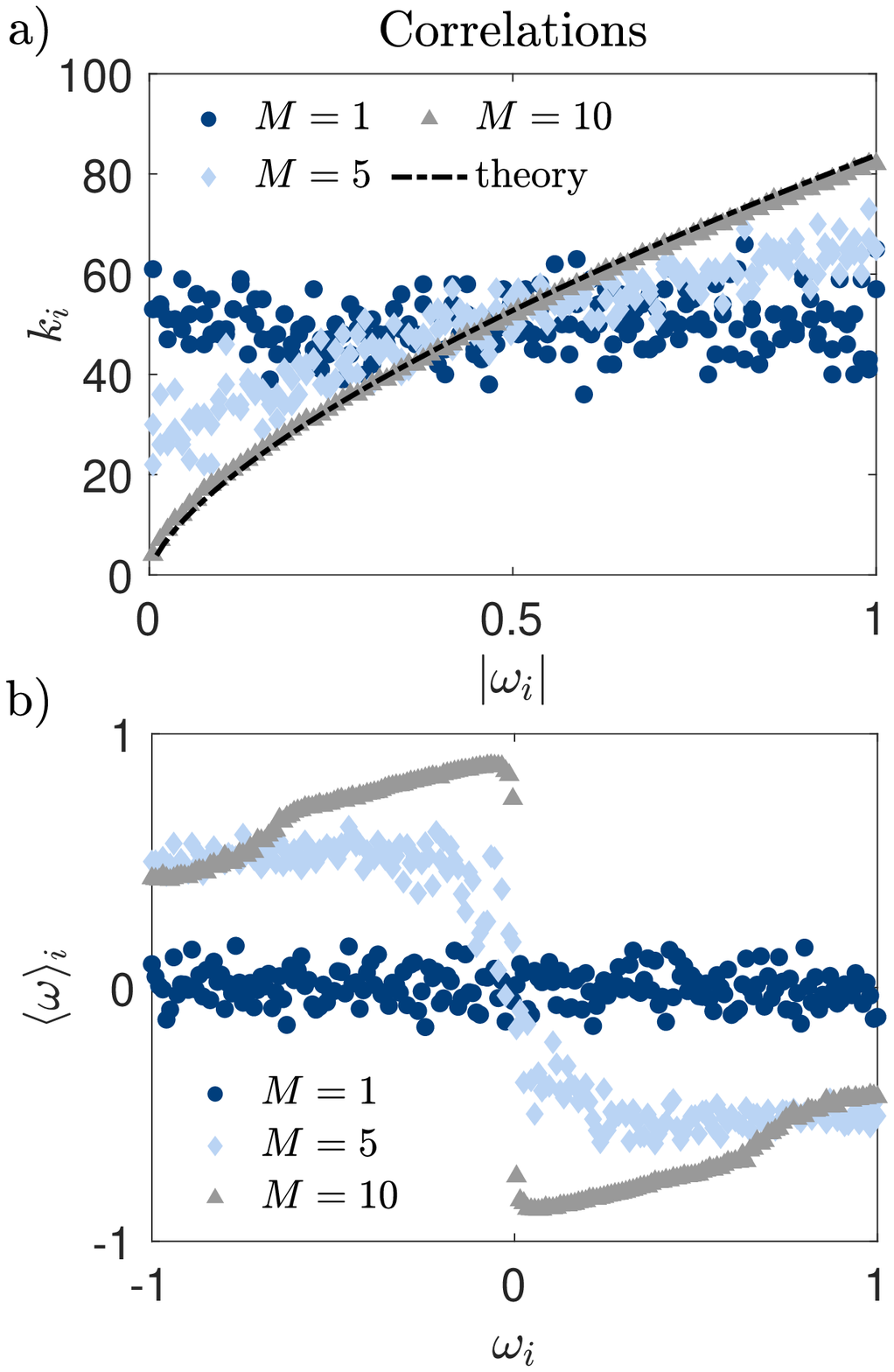}
\includegraphics[scale=0.35]{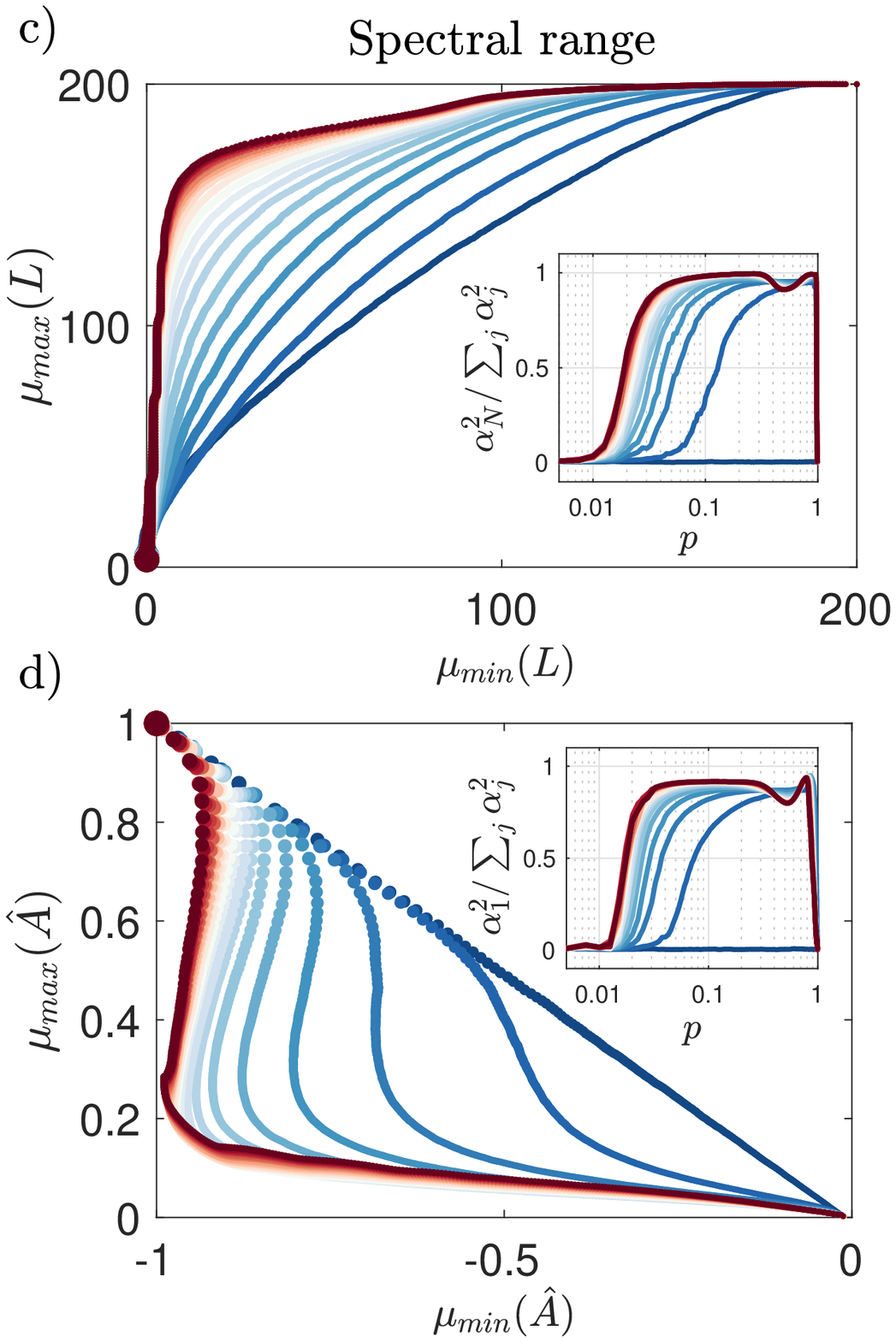}
\includegraphics[scale=0.35]{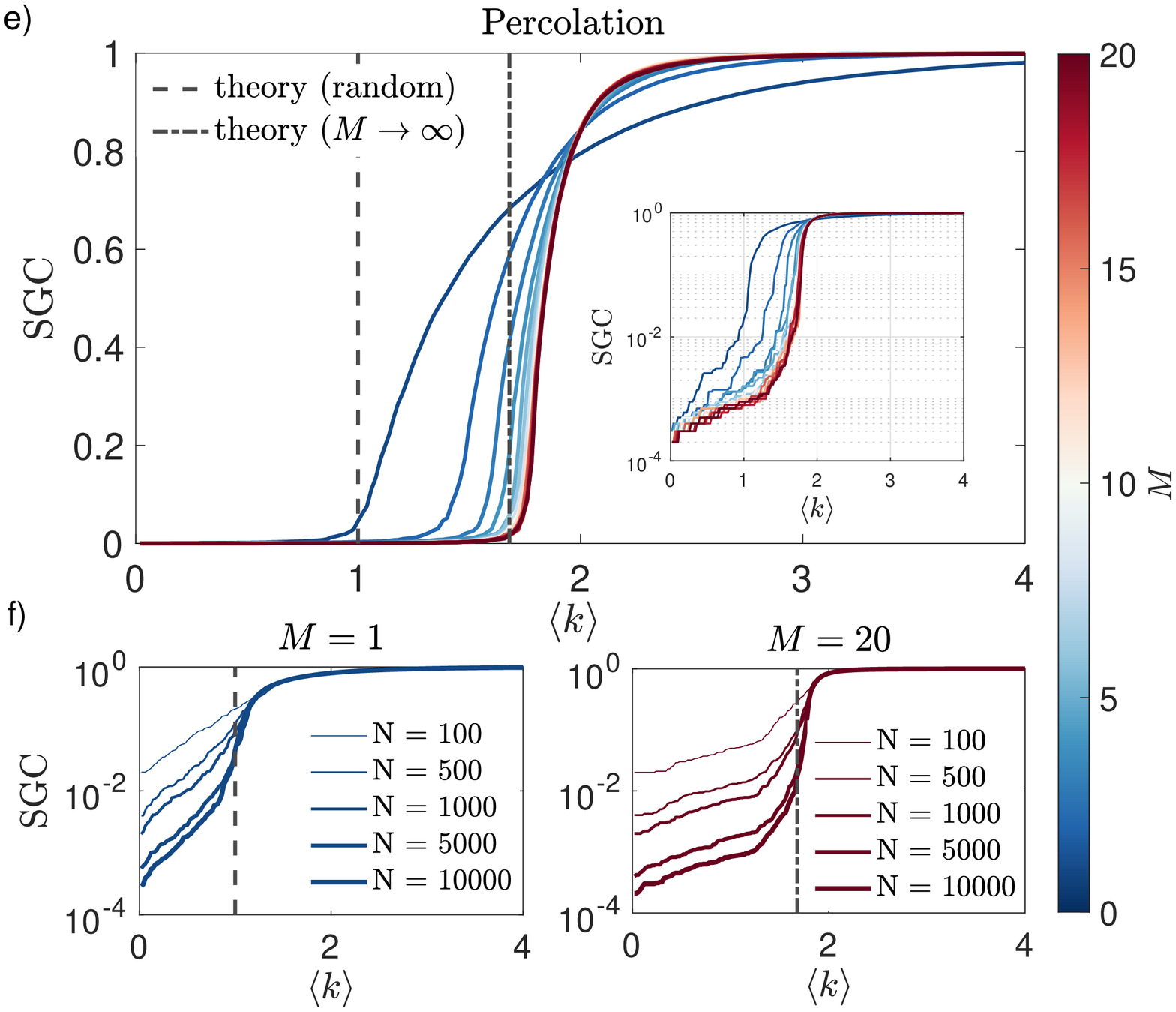}
\caption{Left column: a) Scatter plot of the tuples $(| \omega_i|,k_i)$ and b) $(|\omega_i|,\langle\omega\rangle_i)$ -where $\langle \omega\rangle_i$ is the average frequency of the neighbours of the $ith$-node- for each oscillator at $p = 0.1$, in a single realization of the forward process, for three values of noise and $N = 200$, including in a) the deterministic prediction of Eq.~(\ref{eq:kwcorr}). Middle column: Scatter plot of (c) the maximum and minimum eigenvalues of the Laplacian matrix, $\mu_{min}(L)$ and $\mu_{max}(L)$ and (d) of the normalized Adjacency matrix, ${\mu}_{min}(\hat{A})$ and ${\mu}_{max}(\hat{A})$ at different $p$-steps of the forward process (the size of the dots decreases for larger $p$) for different values of $M$. In the inset, we plot the relative correlation $\alpha_i = \langle \omega, v_i \rangle$  between the frequency vector and the eigenvector of $L$ (or $\hat{A}$) associated with the maximum (or minimum) eigenvalue. Right column: e) Evolution  of  the average size  of  giant  component (SGC), for noise/sampling ranging from the random scenario of $(M = 1)$ to a more deterministic one with $(M = 20)$ in a network of size $N = 5000$. Results are averaged over $20$ realizations of the  process. Inset in e) shows a single realization in log scale. In f) we show the effect of network size on the percolation transition for the $M=1$ (left) and $M=20$ (right) scenarios. It is observed that larger and more deterministic networks under the rule of Eq.~(\ref{eq:rule1-cont}) experiment sharper transitions.}

\label{fig:structure} 
\end{figure*}

\textbf{\label{subsec:fingerprints}Structural explosive fingerprints.}
Before characterizing the synchronization transition of explosive bombs in more depth, we now focus on the structural changes that the network undergoes during the percolation process governed by Eq.~(\ref{eq:rule1-cont}). In the following we analyze the emergence of several structural and dynamical patterns that are usually associated with explosive transitions \cite{dsouza19,arola21}.

\emph{\label{subsubsect:correltions} Degree-frequency correlations and frequency-frequency anticorrelations.}
During the network growth process, the system tends to a stationary degree distribution (which scales with system density) as noise is reduced in the process (for larger sampling $M$). To understand this effect, we recall that the rule of Eq. (\ref{eq:rule1-cont}) tends to connect pairs of oscillators with large frequency differences and low degrees. When a link is chosen, the degrees of the adjacent nodes increase, reducing the value of $\Delta r_{ij}$ for other potential links of these nodes. This constant competition between fixed frequencies and evolving degrees acts as a self-organized feedback that tends to homogenize the distribution of $\Delta r_{ij}$ among the potential links, and frequencies and degrees become balanced in the precise way that makes $\Delta r_{ij}$ more similar among these --still absent-- links. Since the rule predicts the scaling $\Delta r \sim \omega^2/k^3$, we find that, in the deterministic (large $M$) regime (see Methods, section \ref{sec:methods} for details) the relation is given by

\begin{equation}
k_i\approx \dfrac{5}{3}pN\gamma^{-2/3}|\omega_i|^{2/3},
\label{eq:kwcorr}
\end{equation}
where the scaling is controlled by the mean degree, expressed in terms of the density of links $p$ and size $N$, since $\langle k \rangle \approx pN$. As expected, Eq. (\ref{eq:kwcorr}) becomes more accurate as noise is reduced in the system, as observed in Fig.~\ref{fig:structure}.a). In Fig.~\ref{fig:structure}.b), we see that frequency anti-correlations among pairs of connected nodes are also present in the system and become stronger for decreasing noise (large sampling $M$). Let us note that these type of correlations are explicitly imposed in the majority of studied mechanisms that induce explosive synchronization \cite{dsouza19,boccaletti16} whereas here emerge from a decentralized optimization of the synchronized state. In the following, we explain how these dynamical anti-correlations translate into structural ones. 

\emph{\label{subsubsec:bipartite}Spectral signatures: towards optimal and bipartite networks.}
We study the evolution of the extreme eigenvalues $\mu_{max}$ and $\mu_{min}$ of the Laplacian matrix $L = D - A$ ($D$ is the diagonal matrix of degrees) and of the normalized Adjacency matrix $\hat{A} = D^{-\frac{1}{2}}AD^{-\frac{1}{2}}$ during the percolation process, for different values of sampling $M$. In the central panels of Fig.~\ref{fig:structure}, we observe that the network evolves in a path that maximizes both the largest positive eigenvalue of $L$, $\mu_{max}(L)$, ranging from zero to $N$, and the largest negative eigenvalue of $\hat{A}$, $\mu_{min}(\hat{A})$, ranging from minus one to zero, when noise is reduced in the process (larger sampling $M$). 
Also, the frequency of the oscillators tends to correlate with the entries of the associated extreme eigenvectors (see insets of both panels). These spectral signatures pinpoint that our model evolve towards optimal and bipartite configurations. First, it is well understood that optimal synchronization is achieved by the alignment of the frequencies with the largest eigenvector of the Laplacian matrix and by increasing the magnitude of the associated eigenvalue $\mu_{max}(L)$ \cite{skardal14}. Second, note that the normalized Adjacency matrix, $\hat{A}$, is a stochastic row sum and its spectra is bounded in $\mu(\hat{A}) \in [-1,1]$, with the largest eigenvalue ${\mu}_{max}(\hat{A}) = 1$ if the network is connected. The remaining of the spectra follows  Wigner's semicircle law for random networks, becoming narrower as the link density increases, and it deviates from the random case in the presence of modules (shifting towards positive eigenvalues) or bipartite-like structures (shifting towards negative eigenvalues) \cite{arola21}. Thus, from Fig.~\ref{fig:structure}.d) we observe that bipartite patterns arise as determinism is increased (larger $M$) and the trajectory of the extreme eigenvalues tuple follows a clear asymmetric path towards the all-to-all ($p=1$) limit. This effect shows that the rule derived in Eq.~(\ref{eq:rule1-cont}) induces negative structural correlations (bipartitivity) as a consequence of the negative dynamical correlations that emerge in terms of natural frequencies, and vice-versa.

\emph{\label{subsubsec:percolation}Delayed percolation threshold.}
From the former results, it is clear that as the percolation process evolves, the network self-organizes its architecture according to well-known explosive patterns. Two important issues are how this synchrony-driven percolation is related to the natural one, i.e. that observed when links are chosen at random, and, as we will cover below, how the emergence of a giant component (the proportion of the nodes connected in the largest cluster of the network) is related to the synchronization onset. To address these issues we study the emergence of the giant component as a function of the control parameter $p$ when the rule of Eq.~(\ref{eq:rule1-cont}) is applied for different values of the sampling parameter $M$. In Fig.~\ref{fig:structure}.e), we observe that the proposed rule delays the percolation threshold with respect to the random case, and it produces more abrupt transitions. 
Looking more closely at the effect of the system parameters on the percolation threshold, in Fig.~\ref{fig:structure}.f) we observe that, when increasing both the size of the system (large $N$) and the determinism in the rule (large $M$), percolation transitions become sharper and occur at higher $p$. Nevertheless, the nature of the transition appears to be continuous (\emph{i.e.} second order) even for large system sizes.
We can obtain a rough approximation for the average value of the percolation threshold by using the well-known Molloy and Reed criterion \cite{molloy95} in the deterministic limit and neglecting the negative structural correlations that the rule induces. Using this criterion and leveraging the emergent degree-frequency correlation we obtain, for a uniform $g(\omega)$, (see Methods \ref{sec:methods}) that the threshold is estimated as  
\begin{equation}
p_c = \frac{42}{25 N}, 
\label{eq:pc_percolation}
\end{equation}
which can be written in terms of the percolation threshold in a random network \cite{newman10} as $p_c \approx 1.68\cdot p_c^{rand}$. In Fig.~\ref{fig:structure}.e) we observe that Eq.~(\ref{eq:pc_percolation}) works quite well for sufficiently large $M$. 
More sophisticated analytical tools, as the recently developed feature-enriched percolation framework \cite{artime21}, could improve the predictions under local rules, such as Eq.~(\ref{eq:rule1-cont}), that exploit information both from the degrees and the frequencies of the units. \\

\textbf{\label{subsec:kuramoto}Analytical characterization of the Kuramoto bomb.} Now we explore, by analytical and numerical means, the dynamical regimes of our system depending on the coupling, $\lambda$, and noise, $M$, values. It is important to remark that, despite the apparent simplicity of Eq.~(\ref{eq:kuramoto}), the Kuramoto Model on complex networks does not have an analytical solution and approximations are required to predict the dynamical behavior using the information contained in $A$ and $\bm{\omega}$ \cite{arenas08,dsouza19}. 

\begin{figure}[!h]
\includegraphics[scale=0.38]{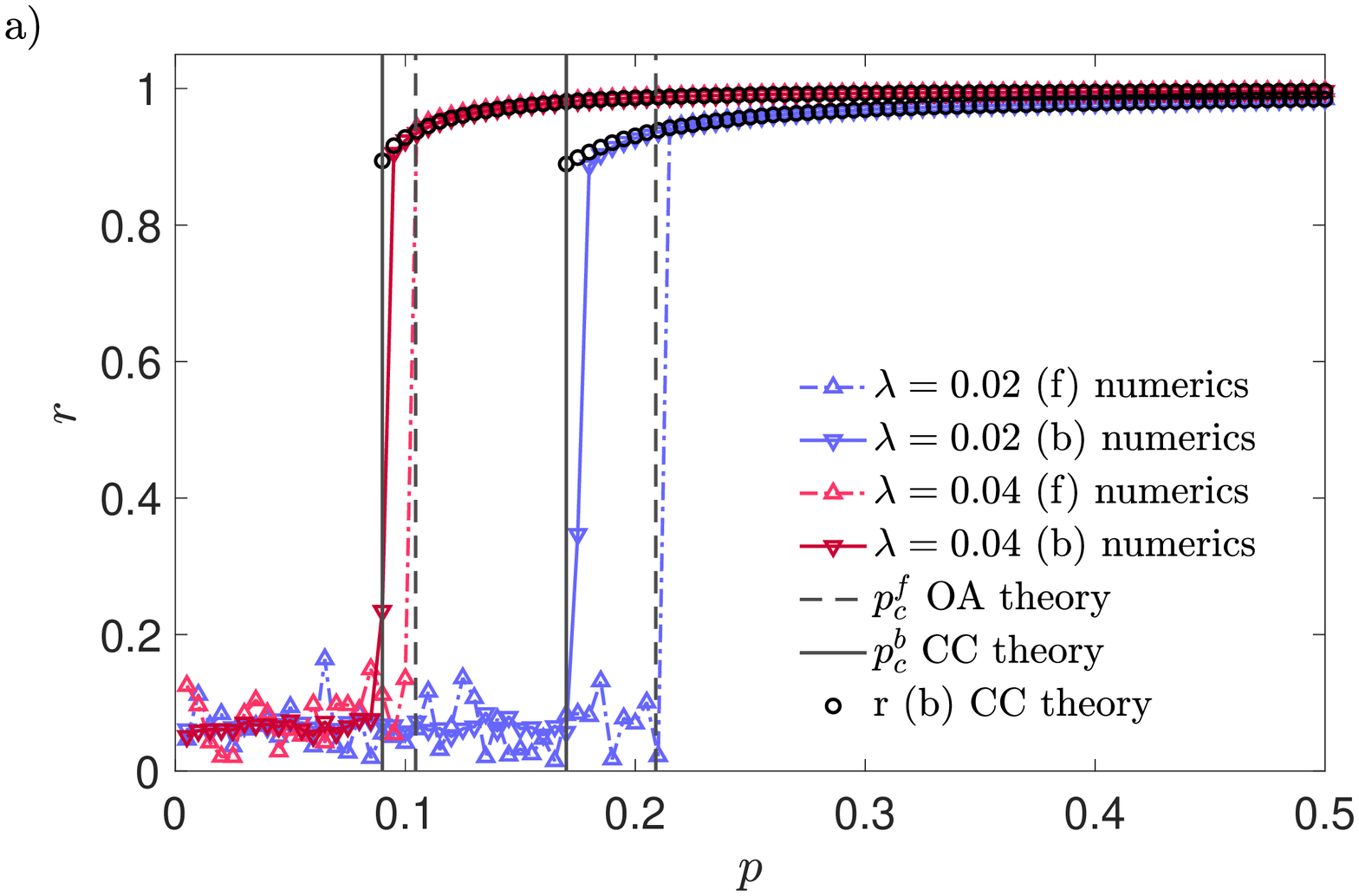}
\includegraphics[scale=0.36]{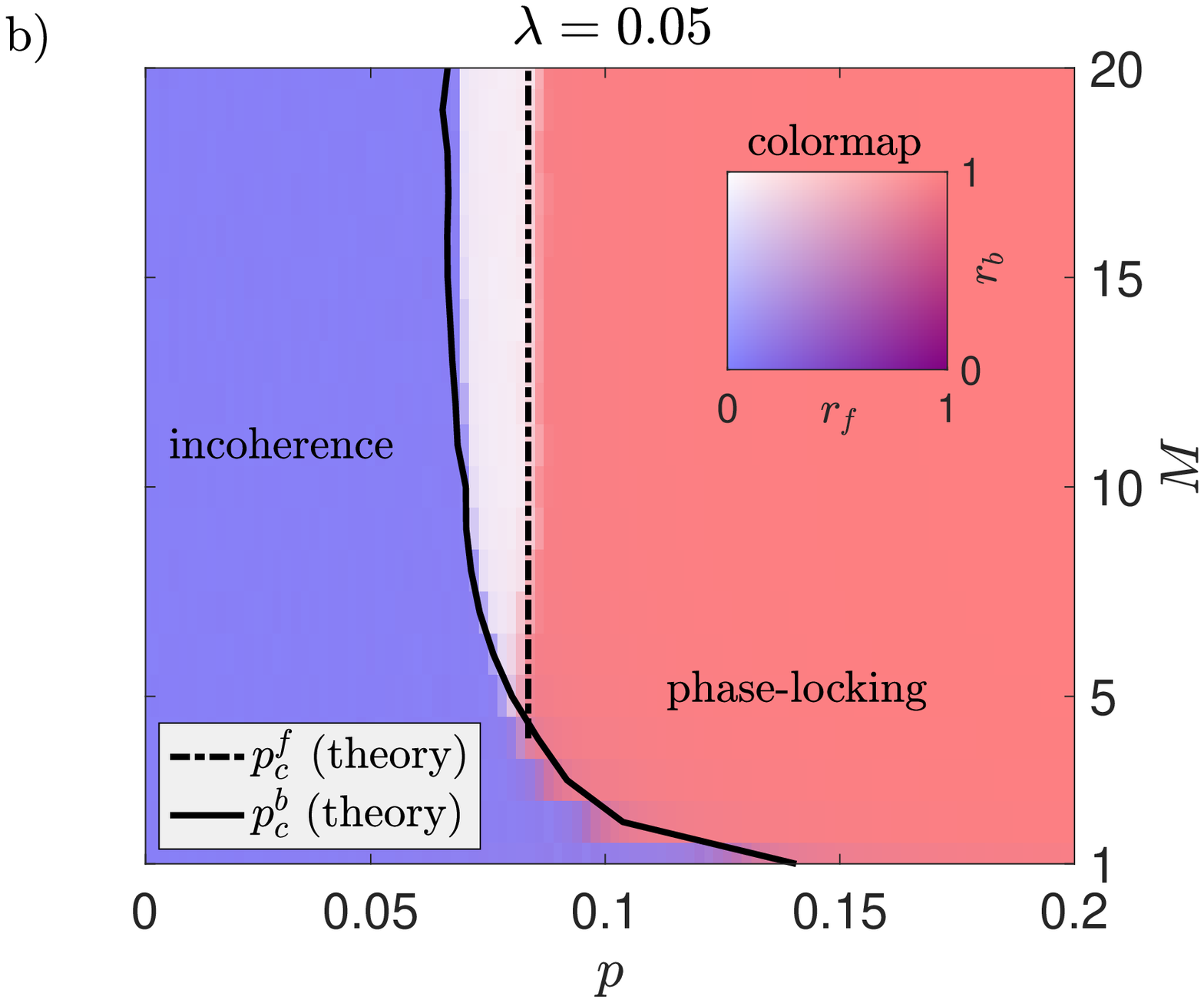}
\includegraphics[scale=0.36]{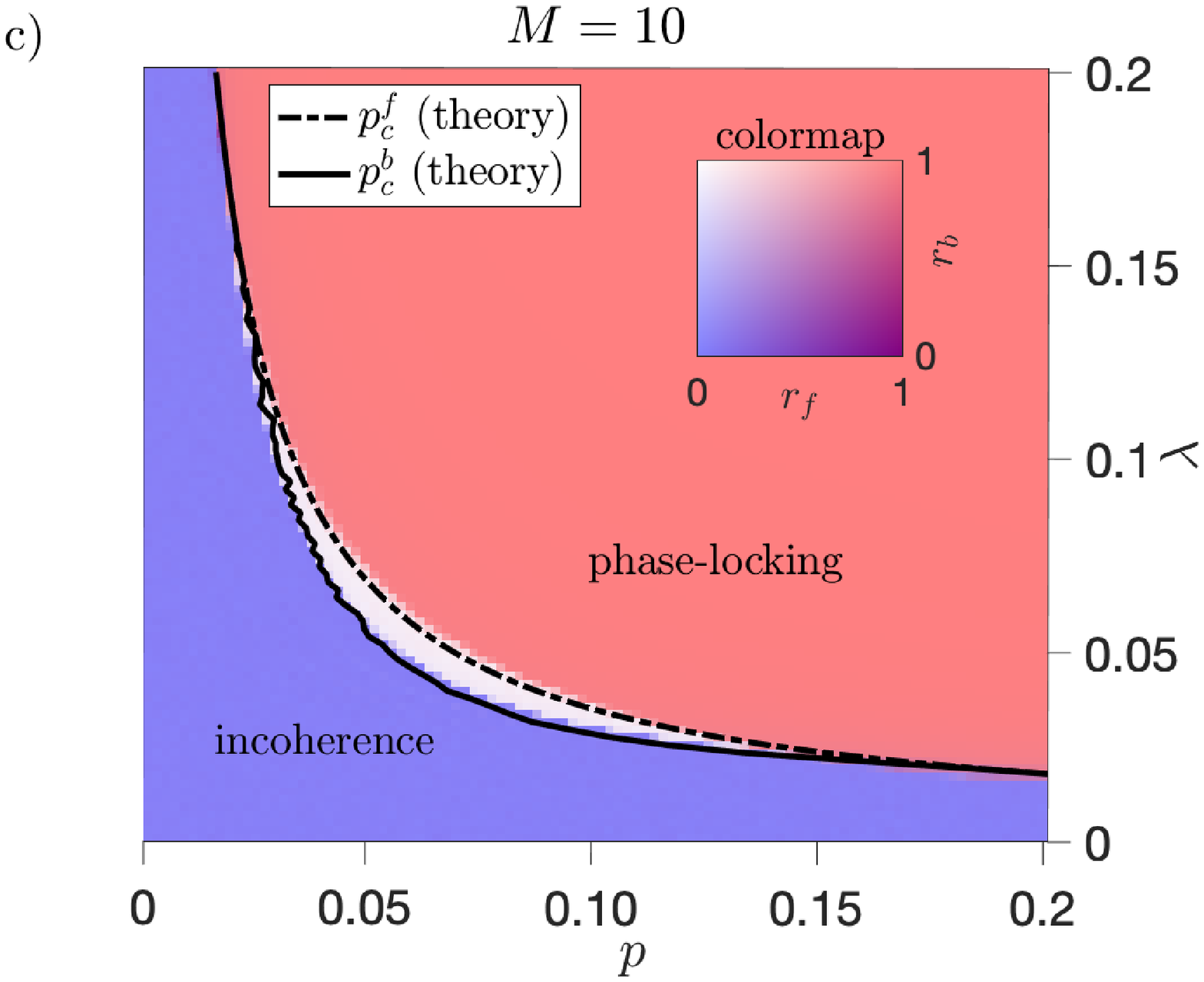}
\caption{a) Two examples of synchronization curves, $r(p)$ in  the forward and backward processes, for a rule dependent $(M =10)$ case in a network of size $N = 200$ and fixed coupling to $\lambda = 0.02$ and $\lambda = 0.04$. Measurements here are taken every 40 links and results are obtained in a single realization of the process. b) Synchronization phase-space depending on $M$ and $p$ for a fixed $\lambda = 0.05$. In the three panels, dashed (solid) lines correspond to the theoretical predictions of the forward (backward) synchronization thresholds, and circle markers in a) give the analytical prediction of the whole backward curve (see main text and \ref{sec:methods} for the derivations). Note that the results displayed in Fig.~\ref{fig:illustrative}.c) correspond to three $M-$slices here. c) Synchronization phase-space depending on $\lambda$ and $p$ for a certain level of noise $M=10$, where the colormap indicates whether the system is in the incoherent, bistable or phase-locking regime. Results are averaged over $20$ realizations.}

\label{fig:XSKuramoto}
\end{figure}

To the best of our knowledge, the current method that better captures the finite-size effects and the precise interplay between the structure and the oscillator dynamics in Eq.~(\ref{eq:kuramoto}) is the model reduction technique based on Collective Coordinates, introduced first by Gottwald to globally coupled systems \cite{gottwald15} and extended to complex networks in \cite{hancock18}. We use this approach to estimate the value of the oscillator phases and the corresponding evolution of the order parameter $r(p)$ in the backward branch and also to calculate numerically the backward synchronization threshold, $p_c^b$. See \ref{sec:methods} for the precise details of this method. The agreement between  CC theory and numerical simulations becomes evident in the backward synchronization diagrams shown in Fig.~\ref{fig:XSKuramoto}.a) for $\lambda=0.02$ and $0.04$ ($M=10$). 

For the forward process we cannot use the CC approach and we rely on the celebrated OA ansatz \cite{ott08}, which has been successfully used to characterize systems in the presence of frequency and degree correlations \cite{restrepo14,skardal15,peron20}. Specifically, we benefit from a recent elegant development used to describe the mean-field dynamics of Janus oscillators \cite{peron20} and consider the limit of large $N$ and $M$. Complete calculations to predict the loss of stability of the incoherent state, and therefore the forward synchronization threshold, are given in the Methods \ref{sec:methods}. For the particular case of a uniform distribution $g(\omega) \in [-1,1]$ we obtain the closed form 
\begin{equation}
p_c^{f} \approx \frac{21}{25\lambda N}.
\label{eq:pc_forward}
\end{equation}
The predicted value $p_c^{f}$ is plotted in Fig.~\ref{fig:XSKuramoto}.a) showing again a remarkable agreeement. This analytical estimation allows addressing the aforementioned issue about the relation between synchronization and percolation onsets by making use of Eq.~(\ref{eq:pc_percolation}) and Eq.~(\ref{eq:pc_forward}). Combining both expressions we can write a simple relation for the percolation $p_c$ and forward synchronization $p_c^f$ thresholds as
\begin{equation}
p_c \approx 2\lambda p_c^f,
\label{eq:pc_relation}
\end{equation}
which illustrates the natural connection between the structural and dynamical aspects of our model.

We extend our numerical and analytical characterization of the synchronization diagram in the $(p,M)$-plane, Fig.~\ref{fig:XSKuramoto}.b), and $(p,\lambda)$-plane, Fig.~\ref{fig:XSKuramoto}.c). In Fig.~\ref{fig:XSKuramoto}.b), we observe that, fixing $\lambda = 0.05$, the collision of the theoretical backward curve and the approximated forward threshold successfully predicts the codimension-two point, where a saddle-node bifurcation collides/appears with a pitchfork bifurcation and bistability emerges \cite{skardal19}. This critical point for which explosive behavior shows up takes place around $M \approx 5$. In Fig.~\ref{fig:XSKuramoto}.c), we focus on the coupling strength $\lambda$, a parameter that does not play a role in the percolation process but it is crucial to synchronization dynamics. The precise location of the synchronization thresholds can be controlled from occurring simultaneously with the percolation one for large values of $\lambda$, to occur much later for smaller values of $\lambda$ and to finally disappear for sufficiently small $\lambda$. Interestingly, the system transits more abruptly for large $\lambda$ (low $p$), but has a larger region of hysteresis for low $\lambda$ (large $p$). 

In the SI, we explore the dynamics of the model for different system sizes, confirming that the abrupt jump in $r$ occurring at single link changes remains large for increasing size, and we show that both phenomenology and theory are robust to changes in the distribution of intrinsic frequencies, $g(\omega)$. In particular, we show results for Gaussian (and bimodal) cases, exploring scenarios with less (and more) polarization than the uniform distribution, finding the expected result that polarization in $g(\omega)$ increases the bistable regime and the abruptness of the transitions. Furthermore, we show that the bomb-like transitions also occur for directed networks, and we analyze in more detail the role of the noisy sampling in the model, finding that an optimal amount of noise can enhance the explosive performance of the synchronization bomb because it improves the self-organized optimization process driven by a local rule.

\textbf{\label{subsec:rossler} Chaotic synchronization bombs}. One of most relevant applications of synchronization theory is its implementation when coupling chaotic systems \cite{BOCCALETTI20021}, a counter--intuitive nonlinear phenomenon as it achieves a perfect dynamical coherence between systems that, when isolated, display exponential divergence of nearby trajectories. Thus, to show the generality of our results, we round off by extending them beyond the Kuramoto framework and considering the R\"ossler system, a paradigmatic model for the emergence of chaotic dynamics  \cite{rossler76}. 

\begin{figure}[!h]
\includegraphics[scale=0.37]{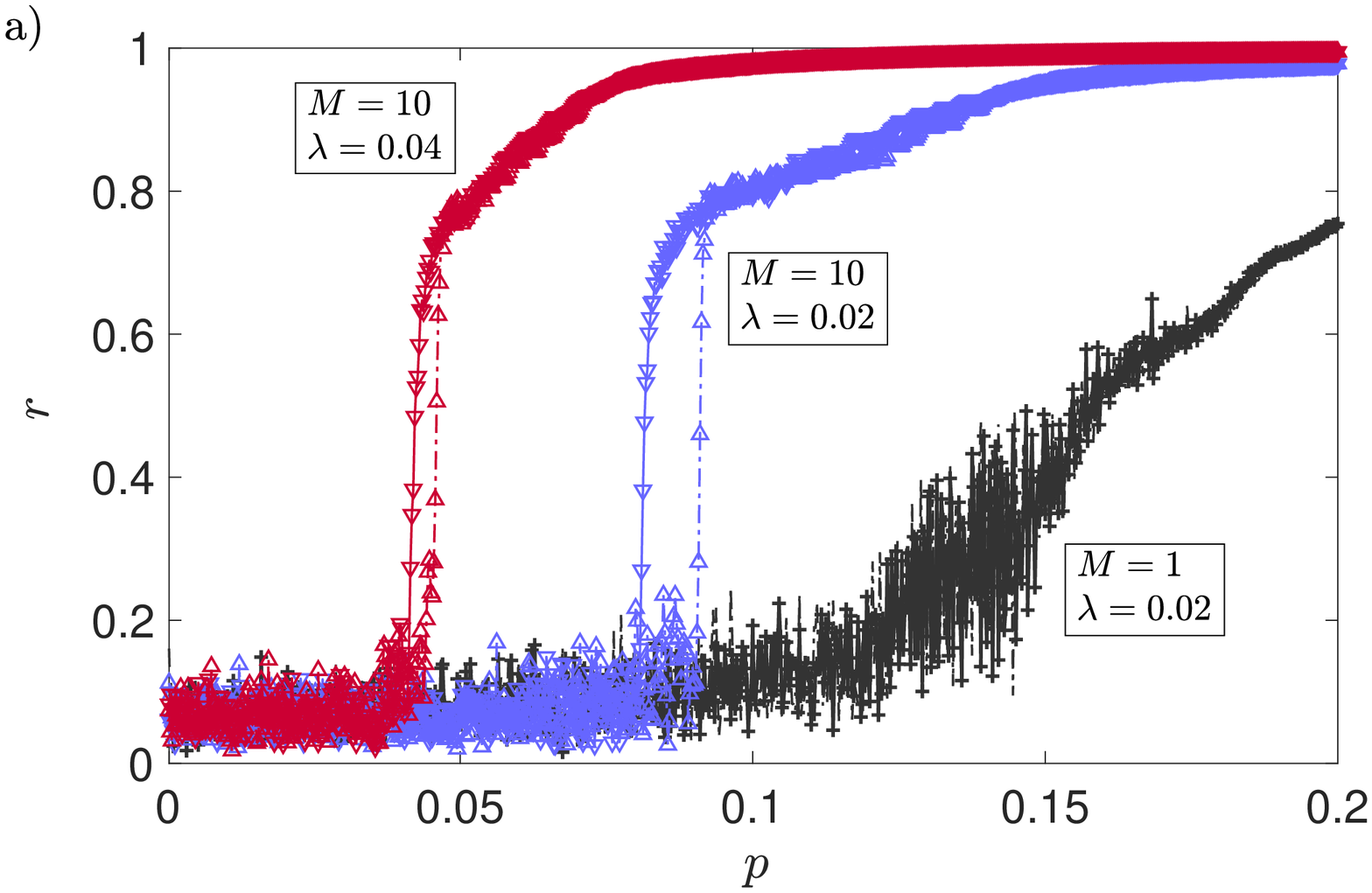}
\includegraphics[scale=0.35]{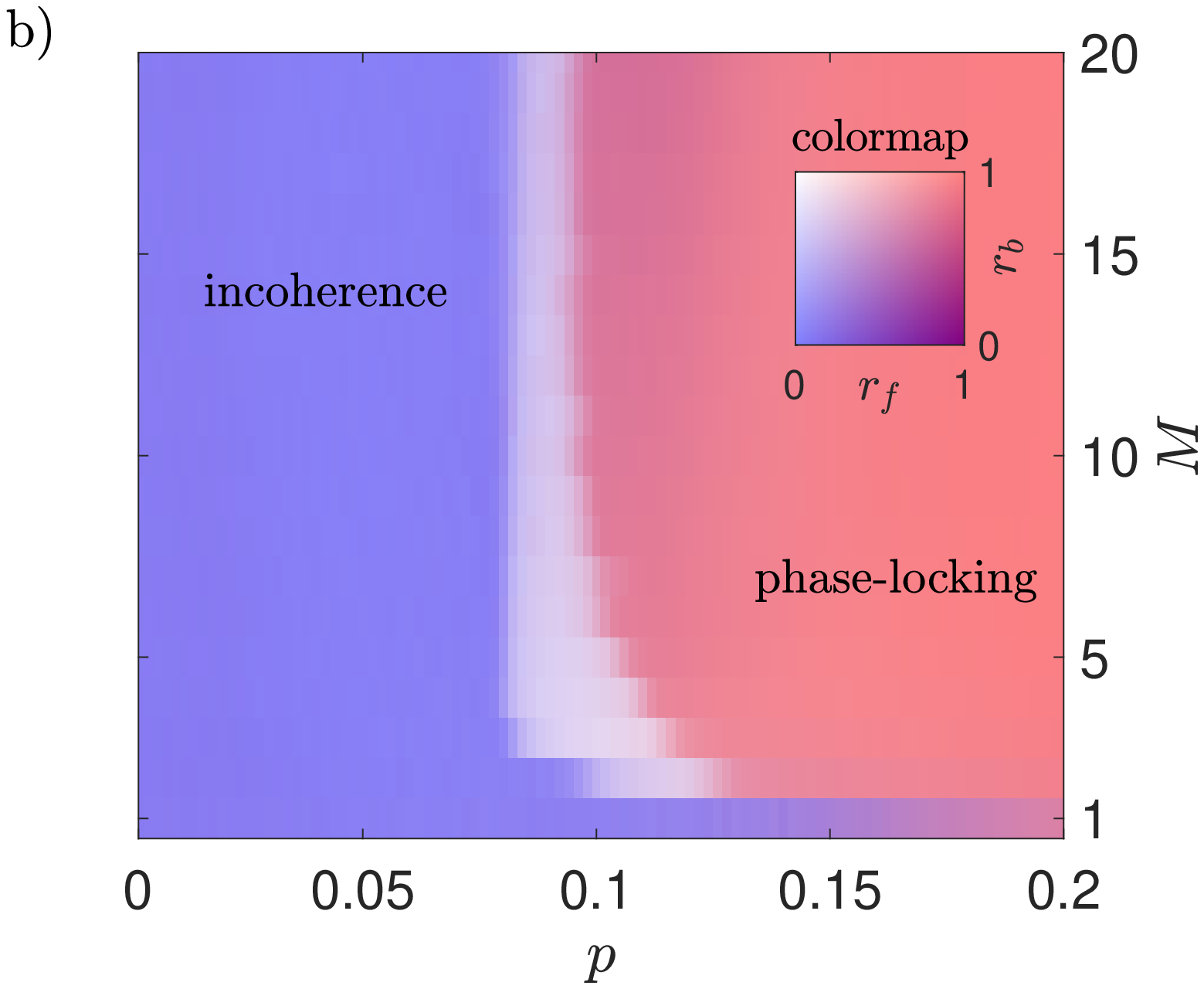}
\includegraphics[scale=0.36]{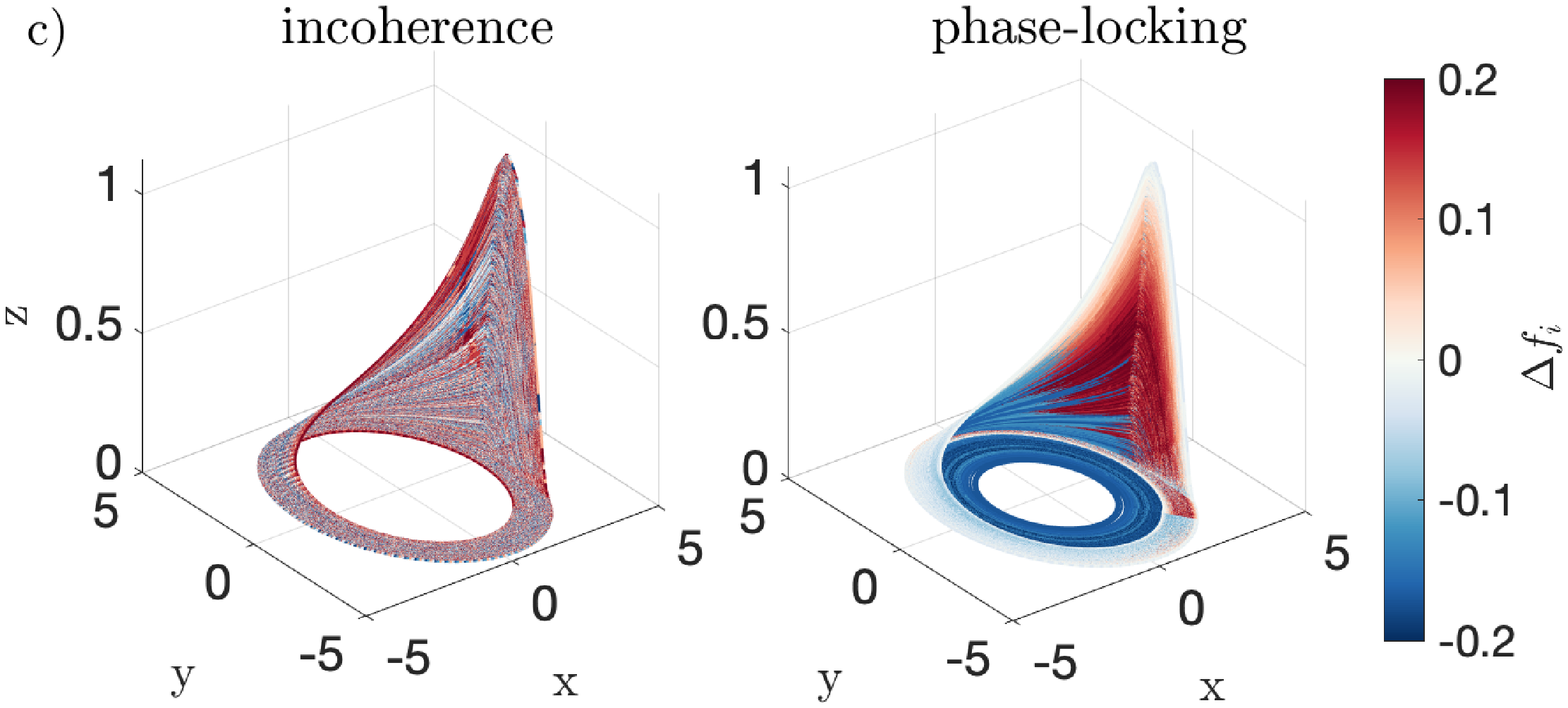}

\caption{a) Example of synchronization curves for several values of $M$ and $\lambda$. It is observed that a hysteresis cycle appears when $M>1$ and that lower (higher) values of $\lambda$ translate into wider (more narrow) cycles and less (more) abrupt transitions. b) Synchronization phase-space, depending on $p$ and $M$ for a fixed $\lambda = 0.02$. Results are qualitatively similar to the ones found in Fig.~\ref{fig:XSKuramoto}.b)), although here the transitions are less abrupt and narrower than in the Kuramoto case, and the birth of hysteresis occurs for higher noise (lower $M$). c) Evolution of the oscillators trajectories in the R\"ossler attractor for $M=10$ and $\lambda = 0.02$ at two different $p$-steps, before (left) and after (right) the forward synchronization transition. The color bar corresponds to the frequency of the oscillator relative to the mean.  Eq.(\ref{eq:rossler-system}) is numerically integrated using Heun's method, with $dt = 0.05$ and $10^3$ time steps and temporal averages of $r$ are taken at every $5$ link changes.}
\label{fig:XSRossler}
\end{figure}

Here we use an ensemble of diffusively coupled heterogeneous chaotic oscillators \cite{roseblum96,leyva12,skardal17}, a modified, piece-wise linear R\"ossler system \cite{rossler76}, which evolves in a 3-dimensional space following
\begin{equation}
\begin{array}{c}
\dot{x}_i = -f_i \left[\tau \left( x_i - \lambda\sum_{j = 1}^N a_{ij}(x_j-x_i)\right)+\beta y_i + \delta z_i \right], \\
\dot{y}_i = -f_i \left(-x_i + \nu y_i \right),\\
\dot{z}_i -f_i \left( -g(x_i)+z_i\right), 
\end{array}
\label{eq:rossler-system}
\end{equation}
where the non-linear function that induces the chaotic behavior is defined as $g(x) = 0$ if $x \leq 3$ and $g(x) = \mu(x-3)$ if $x > 3$. The remaining parameters are set following \cite{leyva12,skardal17}, with $\tau = 0.05$, $\beta = 0.5$, $\delta = 1$, $\nu = 0.02-100/R$. $R = 100$ ensures that the system is in a phase-coherent regime \cite{roseblum96,leyva12,skardal17}, where a phase can be defined after projecting onto the xy-plane, i.e. $\theta_i = arctan(y_i/x_i)$, such that the synchronization order parameter $r$ can be measured by the standard Eq.~(\ref{eq:r}). See Fig.~\ref{fig:XSRossler}.c) for a 3D representation of the trajectories of the chaotic, phase-coherent, oscillators at two different $p$-steps of the forward process. As in Eq.~(\ref{eq:kuramoto}), $\lambda$ is the fixed coupling strength and the entries $a_{ij}$ of the adjacency matrix $A$ capture the presence of undirected and symmetric interactions between the oscillators and evolve under the rule of Eq.~(\ref{eq:rule1-cont}). The instantaneous velocity of each unit is determined by $f_i$, which we assign proportional to the frequency, $f_i = 10 + 0.2\omega_i$, drawn again from a uniform distribution $g(\omega)$ in $[-1,1]$. 

Figure~\ref{fig:XSRossler}.a) illustrates three examples of synchronization transitions $r(p)$ for a system of $N = 200$ and different choices of $\lambda$ and $M$. Similarly to the Kuramoto case, it is observed how, in the construction process, for noise values of $M > 1$ the order parameter experiments abrupt jumps from dynamical incoherence of $r\sim 0.1$ to a more coherent state with $r\simeq 0.7$, that continues to continuously grow to stronger synchronization ($r\gtrsim 0.9$) as the link fraction, $p$, increases. For the backward transition the inverse process takes place but with the jump to incoherence happening for lower values of $p$, resulting in a small hysteresis cycle. In panel~\ref{fig:XSRossler}.c) we show the synchronization diagram in the $(p,M)$-plane, where it becomes clear that the bistable region shows up even for very small values of $M$. 

The success of the chaotic synchronization bomb is grounded on previous research that exploits optimal \cite{skardal17} and explosive \cite{leyva12} synchronization properties of the Kuramoto Model on the diffusively coupled R\"ossler system. However, as numerical results in Fig.~\ref{fig:XSRossler}.a)-b) manifest, the phenomenology is slightly noisier than in the Kuramoto case and the tuning of more parameters along with the chaotic behavior of the units may difficult its design and control. From a practical standpoint, these results show that synchronization bombs could be potentially implemented in the lab, at least by means of electronic circuits \cite{leyva12}. \\

\textbf{\label{subsec:cardiac} Application to cardiac pacemaker cells.} 

Lastly, we demonstrate the existence of self-organized explosive synchronization via synchronization bombs in the biologically-plausible application of cardiac pacemaker cells --the collection of cells responsible for generating a strong, coherent pulse that propagates through the entire heart and initiates each contraction \cite{djabella07}--. For simplicity we consider a system of network-coupled pacemaker cells using, for each pacemaker, a two-variable system describing the dimensionless trans-membrane voltage $v$ and gating variable $h$ which summarizes ionic concentrations~\cite{djabella07}. For a system of $N$ such pacemakers the equations of motion are given by
\begin{align}
\dot{v}_i&=\tau_i^{-1}f(v_i,h_i)+K_v\sum_{j=1}^Na_{ij}(v_j-v_i),\label{cardiac01}\\
\dot{h}_i&=\tau_i^{-1}g(v_i,h_i)+K_h\sum_{j=1}^Na_{ij}(h_j-h_i),\label{cardiac02}
\end{align}
where the local dynamics of each $v_i$ and $h_i$ are described by
\begin{align}
f(v,h)&=\frac{h(v+0.2)^2(1-v)}{0.3}-\frac{v}{6},\label{eq:cardiac03}\\
g(v,h)&=\frac{1}{150}+(8.333\times10^{-4})[1-\text{sgn}(v-0.13)]\nonumber\\&~~~~~~~~~~~~\times\{0.5[1-\text{sgn}(v-0.13)]-h\}. \label{eq:cardiac04}
\end{align}
The timescales $\tau_i$ represent local heterogeneity between the different pacemakers, scaling the period of each isolated cell, ultimately resulting in an effective natural frequency for each pacemaker proportional to $\tau_i^{-1}$. Taking a system of $N=200$ pacemakers with  $\tau_i^{-1}$ uniformly distributed in $[0.4,1.6]$ and using coupling strengths $K_v=0.009$ and $K_h=0.0044$ (to indicate a stronger coupling via the voltage diffusion compared to ionic diffusion) we implement the coupled percolation and synchronization dynamics as presented previously in this work. 

\begin{figure}[!h]
\includegraphics[width=0.85\linewidth]{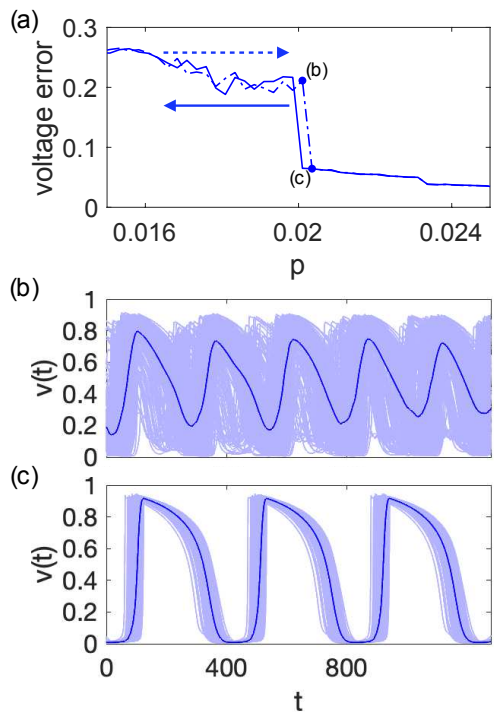}

\caption{(a) Voltage error (quantified by the standard deviation) as a function of the percolation parameter $p$, under forward and backward percolation dynamics plotted in dot-dashed and solid lines, respectively. (b), (c) Individual voltage time series $v_i(t)$ and the mean, plotted in light and dark blue stokes, respectively, from right before and after the explosive transition, as indicated in panel (a).}
\label{fig:cardiac}
\end{figure}

To measure the synchronization of the full system we consider the error in the voltage dynamics, quantified by the overall standard deviation. Taking temporal means of the error as the percolation dynamics are run forward and backwards, we plot the voltage error in Fig.~\ref{fig:cardiac}(a). Note that at roughly $p\approx0.02$ the system undergoes an explosive transition from (relatively) large to small errors, indicating weak and strong synchronization. In Figs.~\ref{fig:cardiac}(b) and (c) we present the actual voltage dynamics right before and after the explosive transition, plotting each individual voltage time series $v_i(t)$ in a light blue stroke, and indicating the overall mean using a thick, dark blue stroke. Note here the physiological implications of the pacemakers ability or inability to produce a strong, coherent pulse for strongly and weakly synchronized behavior, respectively. Lastly, note that, as with the Kuramoto and R\"ossler dynamics, we observe a region of bistability where the forward and backward percolation result in weak and strong synchronization dynamics, respectively.

\section{\label{sec:discussions}Discussion}
Abrupt and explosive phenomena in the structure and dynamics of complex networks have been one of the most studied phenomena in non-equilibrium statistical physics and nonlinear dynamics in recent years. Not only do they allow us to further our theoretical understanding of phase transitions, but also to develop models that are able to explain and reproduce the changes in the topology and behavior observed in natural and engineered systems, such as biological switches, brain activity and blackouts in power-grids. Motivated by the wide range of applications, network percolation and collective synchronization have become paradigmatic frameworks to understand the explosive changes in the structural and dynamical macroscopic properties of large complex systems. A crucial feature of explosive percolation is that it is induced by applying small localized structural perturbations to the system (addition or removal of a few links) by means of competitive rules that delay the formation of a connected component. This aspect was not explored in the synchronization counterpart, where explosive transitions were usually studied by fine-tuning of global coupling parameters in fixed or evolving structures. Furthermore, while the specific theoretical requirements for the explosive behavior become better understood, there is less knowledge on the actual routes that real systems may follow to self-organize towards these particular configurations. 

In this work, we have attempted to bridge these gaps by deriving a local percolation rule for systems of heterogeneous phase-oscillators under the minimal assumption of maximizing global synchronization with decentralized information and noise. We have shown that under this percolation rule the system behaves as a synchronization bomb. This way the network undergoes an explosive synchronization transition at some point of the wiring process, abruptly switching from incoherence to global phase-locking, and display an hysteresis cycle. We have also shown that as the network grows, it self-organizes in a way that several well-known explosive properties on the network structure show up. Importantly, this growth delays the percolation threshold as compared to the usual random case. We have provided an analytical characterization of the system using state-of-art model reduction techniques, obtaining a fair agreement with numerics and being able to reproduce the bistable region in the synchronization phase diagrams. As we show in the SI, all these results are robust under the variation of model assumptions and parameters, and also hold for directed networks. Interestingly, we find that a noisy, low sampling is beneficial in our model because it improves the decentralized optimization of synchrony driven by the proposed local rule. Finally, we have shown that synchronization bombs can be also obtained for systems of coupled chaotic units, paving the way to their implementation in the lab and in a model of cardiac pacemaker cells, proving potential applications in biological systems.

In a nutshell, our findings show that growing networks of heterogeneous dynamical units can develop to operate in a bistable regime, forming networked switches that display the dynamical-structural correlations that are observed when graphs are tuned to display explosive behavior. Also, engineered networks can be designed to be at the onset of total synchrony in which they show no dynamical coherence but, after a minimal wiring (just one or few links), experience synchronization explosions. This finding provides a justification for naming these systems as synchronization bombs. While the current results provide a self-organized and stochastic route to the emergence of these bombs, alternative, deterministic approaches could lead to a better optimization of the explosive behavior and control of the location of the transitions in empirical networked systems.  From a theoretical perspective, triggering the bomb by means of a single local rule and imposing localized, instead of global, perturbations in the system, deepens the explosive connection between synchronization and competitive percolation \cite{zhang14,dsouza19}, and it provides a missing explanation for the birth of abrupt synchronization in pair-wise networks via a universal route \cite{kuehn21}. By switching on a single additional parameter (the amount of sampling in the percolation process), an oscillator network can self-organize towards a high-dimensional correlated state where explosive behavior spontaneously emerges.

\section{\label{sec:methods}Methods}

\textbf{\label{subsec:rule}Derivation of the local rule.}
We begin with two key assumptions: \emph{i)} the system  attempts to maximize the overall degree of synchronization, by adding or removing undirected connections in a percolation process and \emph{ii)} only limited information is available, making this percolation a decentralized process. This means that the units have access only to their immediate surroundings and they can exploit only local information to maximize synchronization, without having access to the overall network synchronization. In order to derive the rule under the previous assumptions, we invoke linearization arguments on the original system Eq. (\ref{eq:kuramoto}), which are shown to be valid when looking for optimal structural and dynamical properties even far away from the linearized regime \cite{dorfler13,skardal14}. Under the linearization, the resulting system reads in matrix form as 
\begin{equation}
\bm{\dot{\theta}} = \bm{\omega} - \lambda L\bm{\theta},
\label{eq:linearized}
\end{equation}
where $L = D - A$ is the Laplacian of the network. The solution of Eq.~(\ref{eq:linearized}) in the stationary state is found by setting $\bm{\dot{\theta}} = 0$. In the corotating frame at speed $\langle w \rangle = 0$, the solution reads as 
\begin{equation}
\bm{\theta}^* \approx \frac{1}{\lambda}{L}^{\dagger} \bm{\omega}, 
\label{eq:statSol}
\end{equation} 
where ${L}^{\dagger}$ is the Moore-Penrose pseudoinverse of the Laplacian matrix, which can be constructed via the spectral decomposition of $L$ for undirected networks (see \cite{skardal14} for more details). Since we are close to the synchronization attractor, the phases in Eq. (\ref{eq:r}) can also be expressed in a Taylor expansion \cite{skardal14}. Invoking again linearization, the order parameter is given by 
\begin{equation}
r \approx 1 - ||\bm{\theta}||^2/2N.
\label{eq:expansion}
\end{equation}
In principle, one needs all the spectral information of the network to estimate the value of $r$. However, we can leverage recent results on the geometric expansion of Eq. (\ref{eq:statSol}) \cite{arola21}, where it is shown that the linearized solution can be expressed as a sum of contributions from increasingly further neighborhoods. This way, the local approximation of synchrony \cite{arola21} is obtained by truncating the expansion at its second term (taking into account the effect of the nearest neighbors of the nodes), leading to
\begin{equation}
r \approx 1 - \frac{1}{2\lambda^2 N} \sum_{i = 1}^N \left(\dfrac{\omega_i+z_i}{k_i}\right)^2.
\label{eq:rApprox1-b}
\end{equation}
where $z_i = \sum_{j = 1}^N a_{ij}(\omega_j/k_j)$ is the contribution of first neighbours. For more details on the accuracy of Eq.~(\ref{eq:rApprox1-b}), see \cite{arola21}. From Eq. (\ref{eq:rApprox1-b}), we can estimate the local impact in the synchrony of adding or removing a single link between oscillators $(p,q)$. Both discrete (considering single link perturbations) and continuous (using derivatives with respect to the degrees and the approximation $\partial z_n/ \partial k_{n'} \approx \pm \delta_{nn'} \omega_{n'} / k_{n'}$, evaluating the resulting expression at $z_i = 0$) calculations, in the limit of large degree, lead to Eq.~(\ref{eq:rule1-cont}) in the results section, an expression that depends only on the local variables $(\omega_i,k_i)$ of a given pair of nodes. Explicitly
\begin{equation}
\Delta r_{ij} = \frac{ \pm 1}{\lambda^2 N} \left( \frac{\omega_i}{k_i} - \frac{\omega_j}{k_j}\right)\left(\frac{\omega_i}{k_i^2} - \frac{\omega_j}{k_j^2}\right),
\label{eq:rule1-cont_METH} 
\end{equation}
where the $\pm$ sign accounts for the addition (removal) of a link. It is important to remark that this result is derived assuming no bias in the coupling function of  Eq.~(\ref{eq:kuramoto}), symmetric and unweighted interactions and a frequency distribution of zero mean, meaning that the actual frequencies of the oscillators may need an appropriate shift to satisfy the condition \cite{arola21}. Also, note that one could obtain more accurate rules for the maximization of $r$ by using the exact result for the phases given by Eq.~(\ref{eq:statSol}) or by including higher-order terms beyond the local approximation, although this increase of accuracy would require to use either spectral (thus global) information or to go beyond the local variables up to second-neighbours and so on. Furthermore, we note that a quadratic approximation of Eq.~(\ref{eq:rule1-cont_METH}) as $\Delta r \sim (\omega_i/k_i - \omega_j/k_j)^2$ also induces the explosive phenomena and may simplify the analytical treatment, but its study is left for further research. \\

\textbf{\label{subsec:percTh}Derivation of the percolation threshold.}
The percolation threshold is approximated by the Molloy and Reed criterion \cite{molloy95}, that predicts the transition for random network without correlations for the value of $p=p_c$ at which 
\begin{equation}
\langle k^2 \rangle (p_c)=2\langle k \rangle (p_c).
\label{eq:MRcond2}
\end{equation}
To compute $\langle k\rangle$ and $\langle k^2\rangle$, we consider a uniform distribution, such that $g(\omega)=1/(2\gamma)$ if $\omega\in [-\gamma,\gamma]$ (the same analysis could be done for any other frequency distribution) and also take into account that, explicitly, we have the general correlation $k_i=c|\omega_i|^{2/3}$ where $c$ is a normalization constant which depends on the network density, $p$, as well as the distribution of natural frequencies, $g(\omega)$. Using that $\langle k\rangle=p(N-1)$, we find that $c=p(N-1)/\langle|\omega|^{2/3}\rangle$, with $\langle|\omega|^{2/3}\rangle=\frac{3}{5}\gamma^{2/3}$. Thus we obtain the correlation
\begin{equation}
k_i\approx\frac{5}{3}pN\gamma^{-2/3}|\omega_i|^{2/3},
\label{eq:kwcorr2}
\end{equation}
and with it
\begin{equation}
\langle k^2\rangle\approx \dfrac{25}{9}p^2 N^2\gamma^{-4/3}\langle|\omega|^{4/3}\rangle = \dfrac{25}{21}p^2 N^2.
\end{equation}
Thus, substituting in Eq. (\ref{eq:MRcond2}) we obtain
\begin{equation}
p_c \approx \dfrac{42}{25}\cdot\dfrac{1}{N} = 1.68\cdot p_c^{rand},
\end{equation}
which corresponds to Eq.~(\ref{eq:pc_percolation}) in the results section. \\


\textbf{\label{subsec:backw}Collective coordinates ansatz}. We use the theory introduced in \cite{gottwald15, hancock18} to predict the phases of the oscillators at any given $p$-step of the backward process and also the transition from phase-locking to incoherence. The main idea of the method is to reduce the dimensionality of the system by considering, as an \emph{ansatz}, that the phases of the oscillators in the phase-locking regime are in the form 

\begin{equation}
    \theta_i = q(t)\psi_i,
    \label{eq:ansatzcc} 
\end{equation}
where $\psi_i$ is the exact solution of the linearized dynamics of Eq.~(1) \cite{skardal14}, i.e. $\psi_i = \frac{1}{\lambda}L^{\dagger}\bm{\omega}$. By minimizing the error made by Eq.~(3) in the full dynamics of Eq.~(1) and after some manipulation \cite{hancock18}, one ends up with only one differential equation for the evolution of the $q$ coefficient, thus drastically reducing the dimensionality from $N$ coupled differential equations to a single one. The resulting equation reads as 
\begin{equation}
\dot{q} = 1 + \frac{1}{\psi^T L \psi}\sum_{i,j}\psi_i \sin(q(\psi_j - \psi_i)). 
\label{eq:alpha_equation}
\end{equation}
Solving the implicit Eq. (\ref{eq:alpha_equation}) for $\dot{q} = 0$ allows estimating the phases of the oscillators in Eq.~(\ref{eq:kuramoto}) beyond the linear regime of the system. Here, we use this theory to predict the phases of the oscillators and the corresponding curve for the order parameter in the full phase-locking regime of the system. Furthermore, to predict the appearance of the (backward) critical threshold $p_c^b$ within this theory, we use an \emph{explosive trick}. We assume beforehand that in the explosive regime of our system, the backward process transits from full phase-locking to complete incoherence. With this idea in mind, we predict the backward threshold by looking at the last values $(p,\lambda)$ for which Eq. (\ref{eq:alpha_equation}) has a solution. Additionally, we check that the solution is linearly stable by numerically computing the eigenvalues of the Jacobian matrix of the full system in Eq.~(\ref{eq:kuramoto}) around the equilibrium solution $q = \hat{q}$. The Jacobian evaluated at the equilibrium point reads as \cite{hancock18}
\begin{gather}
    J_{ij} = a_{ij}\cos(\hat{q}(\psi_j - \psi_i)), \ i \neq j \nonumber \\
    J_{ij} = -\sum_{k}a_{ik} \cos(\hat{q}(\psi_j - \psi_i)), \ i = j. 
    \label{eq:jacobiancc}
\end{gather}
The system is stable if all the eigenvalues of J are negative. Thus, the backward critical threshold occurs at the last value of $p_c^b$ at which Eq. (\ref{eq:alpha_equation}) admits a solution that is linearly stable. The explosive trick is particularly useful to simplify the calculation because, when considering transitions from full phase-locking to incoherence, we do not need to compute partial synchronized solution involving clusters of smaller size than the whole network \cite{hancock18}. In other words, we predict the loss of stability of the full phase-locking state, which in the explosive regime of our system corresponds to the desired backward synchronization threshold. \\

\textbf{\label{subsec:forw}Ott-Antonsen ansatz}. 
In the forward direction, we cannot use the collective coordinates approach anymore because our system departs from the incoherent state where the ansatz Eq.~(\ref{eq:ansatzcc}) is not valid. Numerical simulations showed that, usually for $M > 1$, the incoherent state $r \approx 0$ remains stable beyond the backward critical transition, thus creating a bistable region and a delayed forward transtion. In order to analytically predict the forward critical threshold, we consider the limit of large $N$ and also large $M$ (towards a deterministic rule). In practice, the following results turn out to be valid even for quite small values such as $N = 200$ and $M = 5$, but it is important to remark that the theory is derived in the infinite size and deterministic limits of the model. Our approach is based on the celebrated OA ansatz \cite{ott08} and follows a very similar development to that shown in \cite{peron20}. 

We begin by defining the local order parameter 
\begin{equation}
    R_i = \sum_j a_{ij}e^{i\theta_j},
\end{equation}
such that Eq.~(\ref{eq:kuramoto}) can be written as 

\begin{equation}
    \dot{\theta}_i= \omega_i + Im[e^{i\theta_i}R_i] \ \forall \ i \in 1, \dots, N. 
    \label{eq:theta_i}
\end{equation}

Following \cite{peron20}, we consider a large ensemble of systems, described by the joint probability density $\rho(\theta,\omega,t)$, with $\theta = (\theta_i,\dots,\theta_N)$ and $\omega = (\omega_i,\dots,\omega_N)$. The evolution of the joint probability has to satisfy the continuity equation \cite{ott08}
\begin{equation}
    \frac{\partial \rho}{\partial t} + \sum_{i = 0}^N \frac{\partial}{\partial \theta_i}(\rho \dot{\theta}_i) = 0. 
\end{equation}
where $\theta_i$ is given by Eq. (\ref{eq:theta_i}). Multiplying the density function $\rho$ by $\prod_{j \neq i} d\omega_j d\theta_j$ and integrating, one obtains the evolution for the marginal oscillator density, $\rho_i(\theta_i,\omega_i,t)$ which reads as \cite{peron20} 
\begin{equation}
    \frac{\partial \rho_i}{\partial t} + \frac{\partial}{\partial \theta_i}(\rho \dot{\theta_i}) = 0. 
    \label{eq:density2}
\end{equation}
Now, the OA ansatz can be applied by expanding $\rho_i$ in a Fourier series and setting the coefficients of the expansion as  $b_{i,n} = \alpha_i^n $ \cite{ott08,peron20}. By inserting the Fourier series with the ansatz in Eq. (\ref{eq:density2}), one ends up with

\begin{gather}
    \dot{\alpha}_i + i \alpha_i \omega_i + \frac{\lambda}{2}(\alpha_i^2R_i - R_i^*) = 0, \ \forall \ i \in 1, \dots, N\\
    R_i = \sum_{j = 1} a_{ij} \int_{-\infty}^{\infty} \alpha_j^*(\omega_j,t)g(\omega) d\omega_j, \ \forall \ i \in 1, \dots, N. 
\end{gather}
where $R^*$ and $\alpha_j^*$ represent the complex conjugate and $i$ the imaginary unit. Now we invoke the large $M$ assumption. In this deterministic limit, the underlying network is purely bipartite, split between nodes with positive frequencies and nodes with negative ones (see the results section and Fig.~\ref{fig:structure}.c)-d)). Also, in this limit, the frequencies of the oscillators are completely determined by their degrees. Then, we can look for solutions $\alpha_i = \alpha_{k,\pm}$ \cite{peron20}, reducing the problem to finding solutions for the coefficients of degree classes in the two groups. The local order parameter in this setting can be written as \cite{peron20}

\begin{equation}
    R_{k,\pm} = \frac{k}{\langle k \rangle}\sum_{k'} k'p_{k'}\alpha_{k',\pm}^*. 
\end{equation}
The frequencies of the degree classes in the two groups are completely determined by the percolation rule for a wide range of $p$, leading to 
\begin{equation}
\omega_{k,\pm} = \pm \left(\frac{k}{c}\right)^{3/2},
\end{equation}
where $c$ is a scaling constant given in Eq.~(\ref{eq:kwcorr2}). After these considerations, the resulting system can be written as 
\begin{equation}
\begin{split}
\dot{\alpha}_{k,+} = & - i \left(\frac{k}{c} \right)^{3/2} \alpha_{k,+} + \\
 & \frac{\lambda k}{2\langle k \rangle}\left[\sum_{k'} k' p_{k'}\alpha_{k',-}-\alpha_{k',+}^2\sum_{k'} k' p_{k'} \alpha_{k',-}^*\right]\\
\end{split}
\end{equation}
\begin{equation}
\begin{split}
\dot{\alpha}_{k,-} = & + i \left(\frac{k}{c} \right)^{3/2} \alpha_{k,-} + \\
 & \frac{\lambda k}{2\langle k \rangle}\left[\sum_{k'} k' p_{k'}\alpha_{k',+}-\alpha_{k',-}^2\sum_{k'} k' p_{k'} \alpha_{k,+}^*\right]. 
\end{split}
\end{equation}

Since we want to evaluate the stability of the incoherent state $\alpha_{k,\pm} = 0$, we linearize the system above and evaluate it around $\alpha_{k,\pm} = \delta \alpha_{k,\pm} \ll 1$. After neglecting smaller terms of order $\delta \alpha^2$, the dependence on the complex conjugates vanish and we up with the following system for each degree class

\begin{gather}
    \delta \dot{\alpha}_{k,+} = -i\left(\frac{k}{c} \right)^{3/2} \delta \alpha_{k,+} + \frac{\lambda k}{2\langle k \rangle} \sum_{k'}k'p_{k'} \delta \alpha_{k',-}\\
    \delta \dot{\alpha}_{k,-} = +i\left(\frac{k}{c} \right)^{3/2} \delta \alpha_{k,-} + \frac{\lambda k}{2\langle k \rangle} \sum_{k'}k'p_{k'} \delta \alpha_{k',+}.\\
\end{gather}
By defining the variables $\delta x = \sum_{k'}k'p_{k'}\delta \alpha_{k',+}$ and $\delta y = \sum_{k'}k'p_{k'}\delta \alpha_{k',-}$, and summing over degree classes (taking into account the degree distribution), we can write
\begin{gather}
    \sum_{k} k p_k \delta \dot{\alpha}_{k,+} = -i\sum_{k} \left(\frac{k}{c} \right)^{3/2} k p_k \delta \alpha_{k,+} + \sum_{k} \frac{\lambda k^2 p_k}{2\langle k \rangle} \delta y \\
    \sum_{k} k p_k \delta \dot{\alpha}_{k,-} = +i\sum_{k} \left(\frac{k}{c} \right)^{3/2} k p_k \delta \alpha_{k,-} + \sum_{k} \frac{\lambda k^2 p_k}{2\langle k \rangle} \delta x. 
\end{gather}
With the approximation $\sum_{k} k^{5/2} p_k \delta \alpha_{k,+} \approx \langle k^{3/2} \rangle \delta x$ and $\sum_{k} k^{5/2} p_k \delta \alpha_{k,-} \approx \langle k^{3/2} \rangle \delta y$, the set of equations reduces to a $2$-dimensional variational system for the evolution of $\delta x$ and $\delta y$ that reads as 
\begin{gather}
\delta \dot{x} = -\frac{i \langle k^{3/2} \rangle}{c^{3/2}} \delta x + \frac{\lambda \langle k^2 \rangle}{2 \langle k \rangle} \delta y \\
\delta \dot{y} = \frac{\lambda \langle k^2 \rangle}{2 \langle k \rangle} \delta x + +\frac{i \langle k^{3/2} \rangle}{c^{3/2}} \delta y
\end{gather}
It is straightforward to show that the critical condition for the stability of the incoherent state is given by 
\begin{equation}
    c^{3/2} \lambda \langle k^2 \rangle = 2\langle k^{3/2} \rangle  \langle k \rangle.
    \label{eq:critical_condition}
\end{equation}
In particular, the eigenvalues of the Jacobian matrix change from being both imaginary to become both real as density increases in the system. In fact, the fully imaginary spectrum predicts the existence of a center attractor, indicating a marginal stability of the incoherent state. Therefore, one might expect stationary oscillations of the order parameter \cite{peron20}. Here we do not observe these oscillations in the forward process. The system is initialized with isolated units (in the incoherent state) and remains there as the network evolves in an adiabatic manner. Fortunately, the forward abrupt transition to phase-locking is well predicted by the critical condition given by Eq.(\ref{eq:critical_condition}), when the eigenvalues become real (one positive and one negative) indicating the appearance of an unstable saddle point. Accordingly, when the condition is achieved in the forward, growth process, the marginal stability of the incoherent state is lost and the system transits to phase-locking.

Using that in the deterministic limit we have that $k_i = c|\omega_i|^{2/3}$, and for a general $g(\omega)$ the constant is given by $c = \langle k \rangle/\langle |\omega|^{2/3}\rangle$, we obtain a general closed form for the forward critical threshold $(p_c,\lambda_c)$ that is given by 

\begin{equation}
    p_c^f = \frac{2 \langle |\omega|^{2/3}\rangle^2 \langle |\omega|\rangle}{\lambda N \langle |\omega |^{4/3}\rangle}.
    \label{eq:pc_percolation_meth}
\end{equation}
For the particular case of a uniform distribution $g(\omega) \in [-\gamma,\gamma]$ we can easily compute the expected moments and, after plugging these results in Eq.~(\ref{eq:pc_percolation_meth}), we end up with the simple formula 

\begin{equation}
    p_c^f = \frac{21\gamma}{25\lambda N}.
\end{equation}
which corresponds to Eq.~(\ref{eq:pc_forward}) in the results section.

\begin{acknowledgments}
  L.A.-F. and A.A. acknowledge the Spanish MINECO (Grant No. PGC2018-094754-B-C2). JGG acknowledges the Spanish MINECO (Grant No. FIS2017-87519-P), the Departamento de Industria e Innovaci\'on del Gobierno de Arag\'on and Fondo Social Europeo through (Grant No. E36-17R FENOL), and Fundaci\'on Ibercaja and Universidad de Zaragoza (Grant No. 224220). A.A. acknowledges financial support from Generalitat de Catalunya (grant No.\ 2017SGR-896), Universitat Rovira i Virgili (grant No.\ 2019PFR-URV-B2-41), Generalitat de Catalunya ICREA Academia, and the James S. McDonnell Foundation (grant \#220020325). This work was supported by MINECO and FEDER funds through Projects No. FIS2017-87519-P, No. FIS2017-90782-REDT (IBERSINC); from grant PID2020-113582GB-I00 funded by MCIN/AEI/10.13039/501100011033; and
  by the Departamento de Industria e Innovación del Gobierno de Aragón y Fondo Social Europeo through Grant No. E36-17R (FENOL). S. F.-L. acknowledges financial support by Gobierno de Aragón through the Grant defined in ORDEN IIU/1408/2018. E.-C. B. acknowledges support from the “Agencia Estatal de Investigación” (Ref. PRE2019-088482), Government of Spain (FIS2020-TRANQI; Severo Ochoa CEX2019-000910-S), Fundació Cellex, Fundació Mir-Puig, and Generalitat de Catalunya (CERCA, AGAUR).
\end{acknowledgments}

\newpage 

\section{Supplementary Information}

In this supplementary section, we study in more detail the robustness of the explosive phenomena in our model against changes in the main parameters. In particular, focusing on the Kuramoto dynamics, we study the synchronization transitions occurring at single links for different system sizes to validate the abrupt nature of the process when localized perturbations are applied to the system. We measure the maximum jump in the order parameter and the area of hysteresis depending on both size and noise (in terms of the sampling parameter $M$), finding that the explosive phenomena is maximized in the large size limit and, unexpectedly, by an optimal amount of noise that compensates the error made by constantly applying a local --decentralized-- percolation rule. We also check that our numerical and analytical results hold for a Gaussian and a bimodal distribution of frequencies (note that all results in the main text were presented for a uniform choice of $g(\omega)$, which simplified calculations). We conclude by showing results for the Kuramoto bomb in directed networks, where we use the modified percolation rule that correctly accounts for directionality in the chosen links while preserving the decentralized nature of the process. Interestingly, we find that the explosive synchronization transitions also occur in the directed case, although with a bistability window that is smaller than in its undirected counterpart.

\subsection{\label{sec:size}Effect of size and noise}

\begin{figure}[!h]
\center{\includegraphics[scale=0.30]{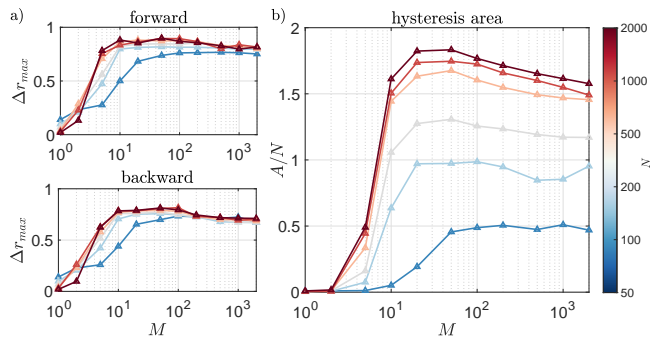}}
\caption{ a) Maximum difference of the average order parameter between two consecutive link changes in the forward (top) and backward (bottom) directions for increasing values of $M$ in log scale and for different sizes (ranging from blue to red). b) Normalized hysteresis area (the sum of differences between the values in the backward and forward curves) for increasing values of $M$ and for different sizes. We observe the monotonous dependence with $N$ and the non-monotonous one with $M$. Results are averaged over $10$ realizations of the process. Each percolation process is run in both directions from $p = 0$ to $p = 20/N$, corresponding to a maximum mean degree $\langle k \rangle = 20$, integrating the KM equations using Heun's method, with $dt = 0.05$ and $10^3$ steps, discarding the first half for averaging $r$. Coupling strength is set to $\lambda = 0.05$ and $g(\omega) \in [-1,1]$.}
\label{fig:SuppSize}
\end{figure}

\begin{figure}[!h]
\center{\includegraphics[scale=0.30]{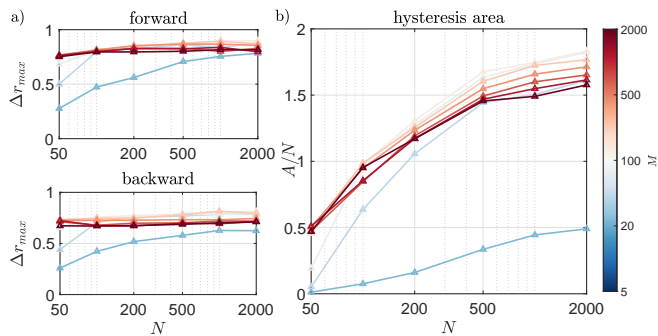}}

\caption{a) Maximum difference of the average order parameter between two consecutive link changes in the forward (top) and backward (bottom) directions for increasing values of size $N$ in log scale and for different sampling or noise $M$ (ranging from blue to red). b) Normalized hysteresis area (the sum of differences between the values in the backward and forward curves) for increasing values of $N$ and for different samplings. We observe again the monotonous dependence with $N$ and the non-monotonous one with $M$. Parametrization is the same as in the previous figure.}
\label{fig:SuppSize2}
\end{figure}

First, we study both the effect of size $N$ and noise (via sampling parameter $M)$ in the synchronization transition emerging from the self-organized network growth. We validate that the abruptness of the transitions occurring at single link changes is sustained for increasing size, such that $\Delta r$ in a single step does not vanish as size grows \cite{dsouza19}. In fact, we observe in Fig. \ref{fig:SuppSize}.a) and Fig. \ref{fig:SuppSize2}.a) that the mean maximum jump value increases monotonically with size (towards red colors) and leads to a macroscopic jump in $r$ at single link changes even for large system sizes. The same occurs in terms of the hysteresis area (normalized by size for proper comparison), which increases for large system sizes, as shown in Fig. \ref{fig:SuppSize}.b) and Fig. \ref{fig:SuppSize2}.b). Interestingly, the dependence on noise, via the sampling parameter $M$ is clearly non-monotonous, showing a peak around $M \approx 50$ regardless of size. 

The counter--intuitive effect of the local rule better sustaining synchronization as noise is widely present (in terms of a low sampling $M$) can be explained by noting that the rule is derived with local information (see the Methods section in the main text), such that higher-order effects are neglected by assumption. However, applying the rule itself makes higher-order effects more important (inducing structural and dynamical anti--correlations). The local prediction of $\Delta r$ for the sampled links may deviate from the exact one as we advance in the percolation process, producing negative feedback that penalizes the maximization of $\Delta r$ as the mechanism becomes more deterministic, and a precise amount of noise leads to the optimal performance. Luckily, the optimal value of $M$ in a particular setting can be estimated by leveraging the analytical results presented in the main text, without running the dynamics. Due to the proven goodness of the CC ansatz in the explosive regime of our system, finding the $M$ that maximizes the degree of synchrony in the linearized solution (which can be directly computed via the aforementioned pseudo-inverse Laplacian) will turn out to be the $M$ that maximizes explosive behavior. However, the reader should note that the proposed model is intrinsically noisy, and the location of the synchronization transitions may vary between different realizations of the process. Note that alternative --deterministic-- methods to build synchronization bombs could minimize this uncertainty, but the current mechanism is intentionally designed in the presence of noise. This noisy aspect turns out to be crucial to the optimal performance of the bombs in our model.

\subsection{\label{sec:CC}Effect of the frequency distribution} 
In the main text we presented our results for a particular choice of the frequency distribution $g(\omega)$ -the uniform one- in order to simplify the analytical treatment. Here we show that our model is robust to different choices of the intrinsic frequencies of the oscillators. In particular, we consider a Normal distribution and a bounded bimodal one, generated with a Beta(0.1,0.1) distribution, a family of continuous probability distributions defined on the interval [0,1], fixing the mean to zero and the variance to $\sigma^2 = 1/3$, in order to compare against the uniform case in [-1,1] used in the main text, which has the aforementioned variance. 
\begin{figure}[!h]
\includegraphics[scale=0.33]{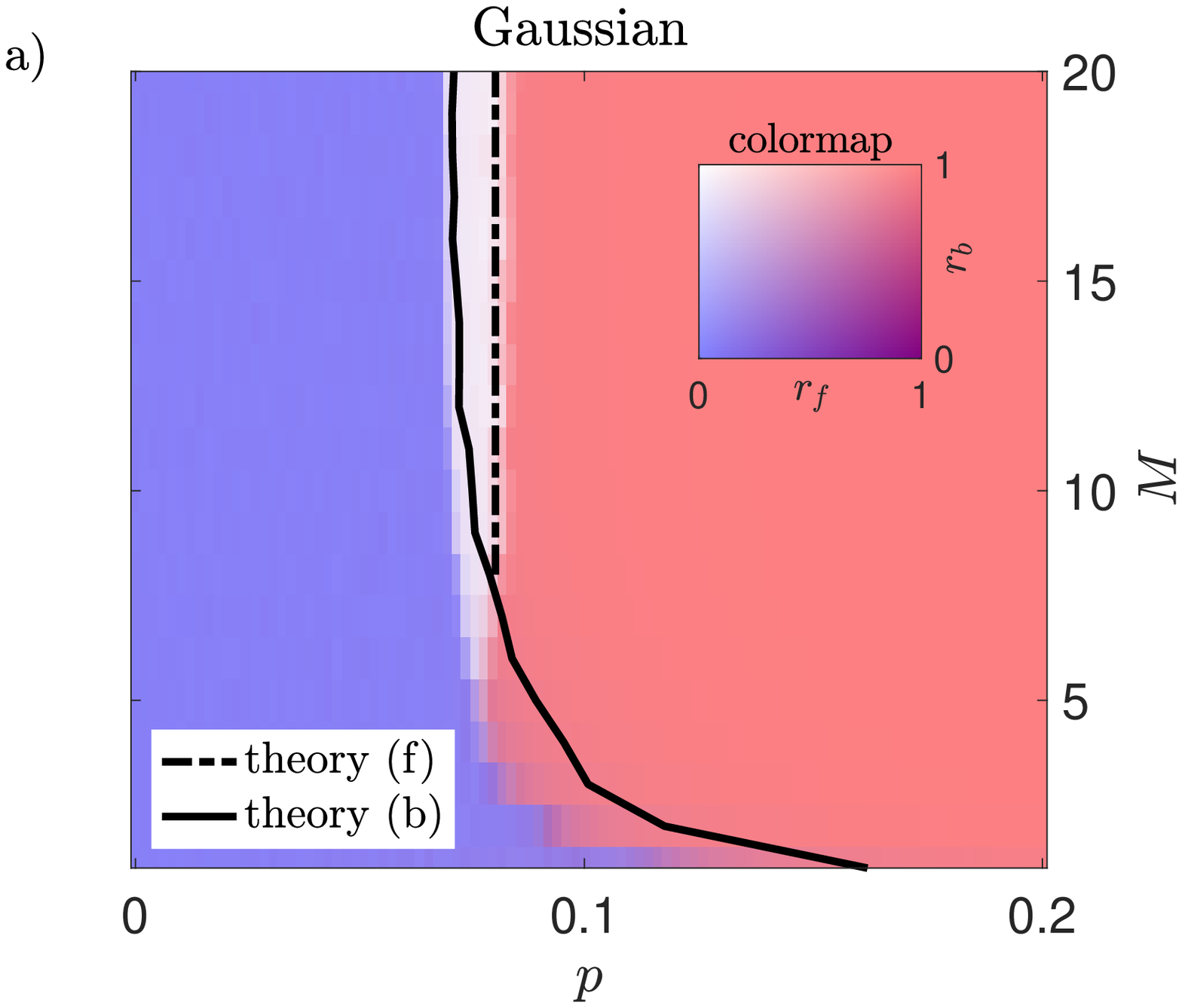}
\center{\includegraphics[scale=0.34]{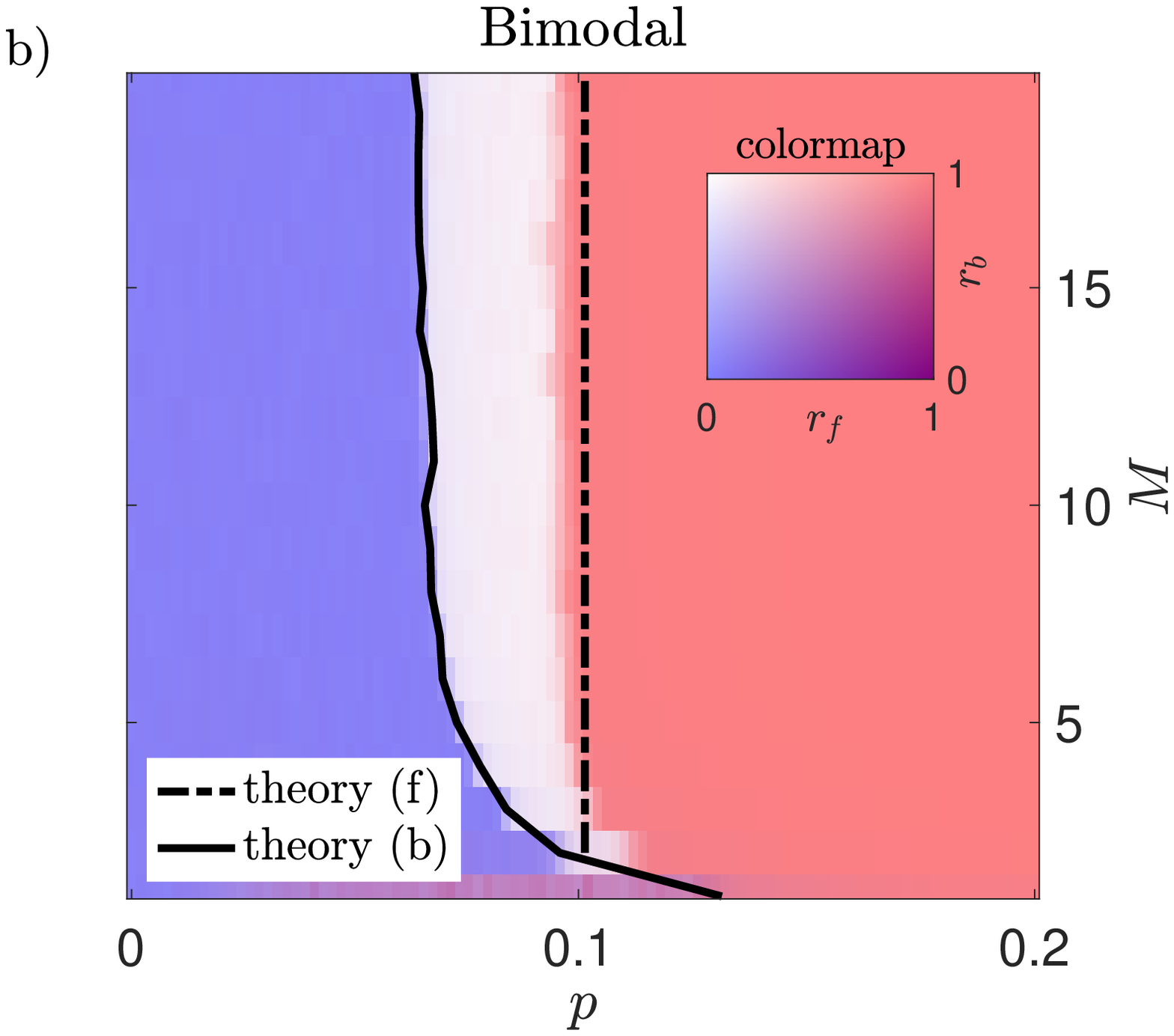}}
\caption{Synchronization phase-space depending on $M$ and $p$ for a fixed $\lambda = 0.05$ for a Gaussian (a) and bimodal (b) distribution of frequencies with mean zero and variance $1/3$. Dashed (solid) lines correspond to the theoretical predictions of the forward (backward) synchronization thresholds (see main text for the derivations).}
\label{fig:SuppFreq}
\end{figure}

In Fig. \ref{fig:SuppFreq} we observe that a clear bistable region in the plane $(p,M)$ also emerges for these choices of $g(\omega)$. The bistable region is larger for the bimodal distribution, which shows that having a more polarized distribution of frequencies enhances the explosivity in the system. In the Gaussian case (less polarized than the uniform one), the bistable region is much narrower, as can be seen in Fig. \ref{fig:SuppFreq}.a). Furthermore, in this case, the prediction of the backward synchronization threshold (solid line) is less accurate than in the other scenarios for low $M$ (high noise). This inaccuracy can be explained by noting that the CC method \cite{gottwald15,hancock18} used to predict the threshold is based on an \emph{explosive trick} that assumes that the whole system is in the phase-locking state before the backward transition. This assumption does not hold for a Gaussian distribution of $g(\omega)$, where the global phase-locking state is not supported by the overall network (just by a large fraction of the oscillators) in the backward process, and the value at which the full phase-locking state loses the stability does not coincide with the backward synchronization threshold. Nevertheless, as discussed in the Methods section in the main text, this theory could be improved by finding the largest synchronized cluster of a given size smaller than $N$, although this improvement demands larger computational costs \cite{hancock18}. The forward prediction (dashed line), based on the OA ansatz \cite{ott08}, does not suffer from this issue and captures well the critical threshold even for high values of noise (low $M$). 

\subsection{Extension to directed networks} 
We close the supplementary information by considering an important extension of our initial results. In particular, in the main text we restricted our model to undirected networks (assuming symmetric interactions, meaning that if $a_{ij} = 1$ then $a_{ji} = 1$). Now we extend these results to a more general setting by allowing directed connections, which may not necessarily be symmetric. 

Convenient to our purpose here, we can leverage the results in \cite{arola21}, which exploited the truncated expansion of the linearized synchronization dynamics in directed networks to predict the existence of links leading to the counterintuitive Braess' paradox in synchronization (a removal of link that increase the degree of synchrony). Following \cite{arola21}, a modified version of our percolation rule (Eq. (3) in the main text) that accounts for the directionality of links and predicts the change of synchrony with local information can be written as 
\begin{equation}
    \Delta r_{ij} = \pm \frac{1}{N k_i} \left[\frac{\omega_p}{k_p}\left(\frac{\omega_j}{k_j}-\frac{\omega_j}{k_j} \right) \right],
    \label{eq:directed-rule} 
\end{equation}
where $\Delta r_{ij}$ accounts for the change in $r$ after adding (or removing) a directed link coming from $j$ to $i$, and $k_i = \sum_j a_{ij}$ is the in-degree of the $i$-node. 

\begin{figure}[!h]
\includegraphics[scale=0.40]{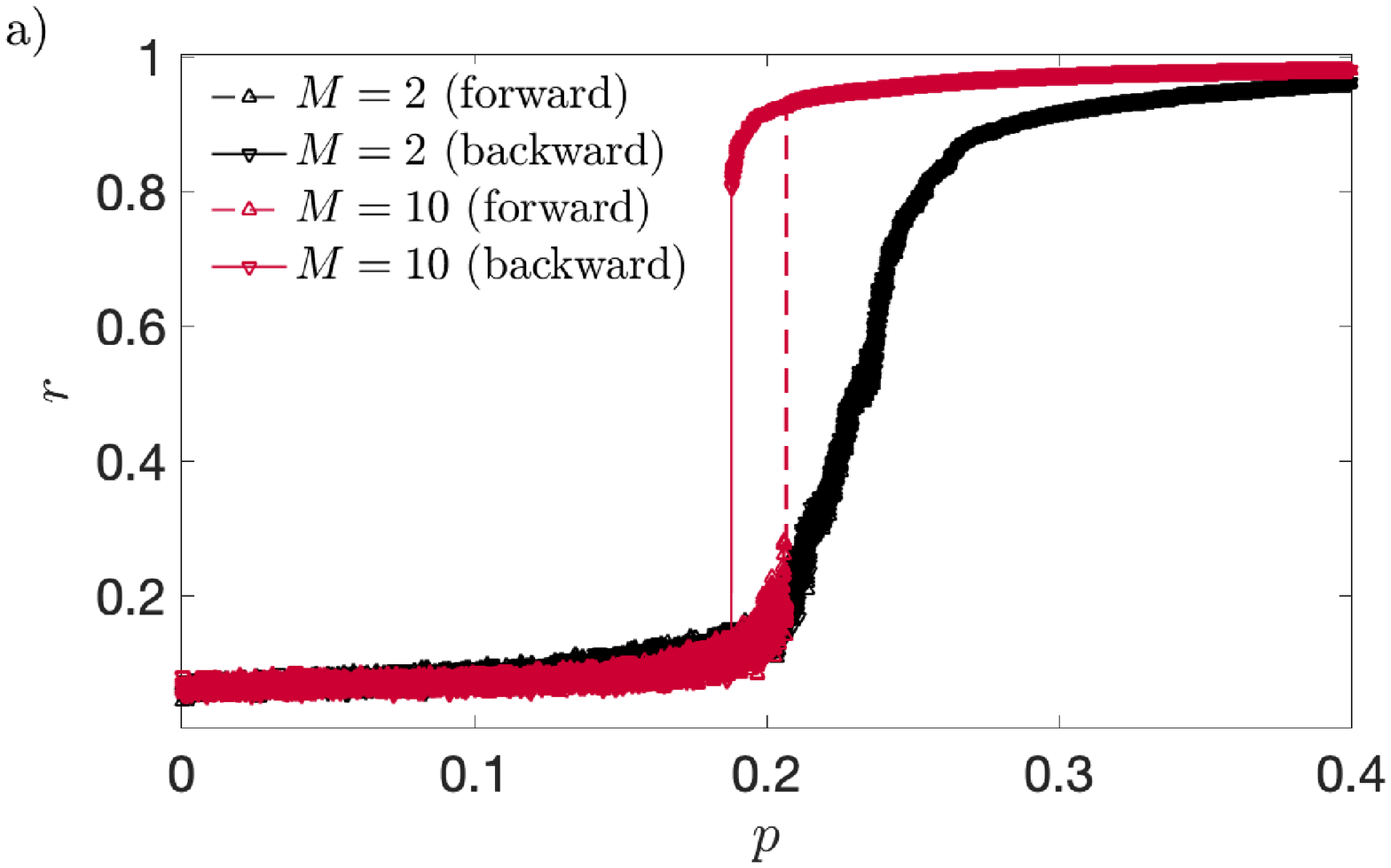}
\center{\includegraphics[scale=0.40]{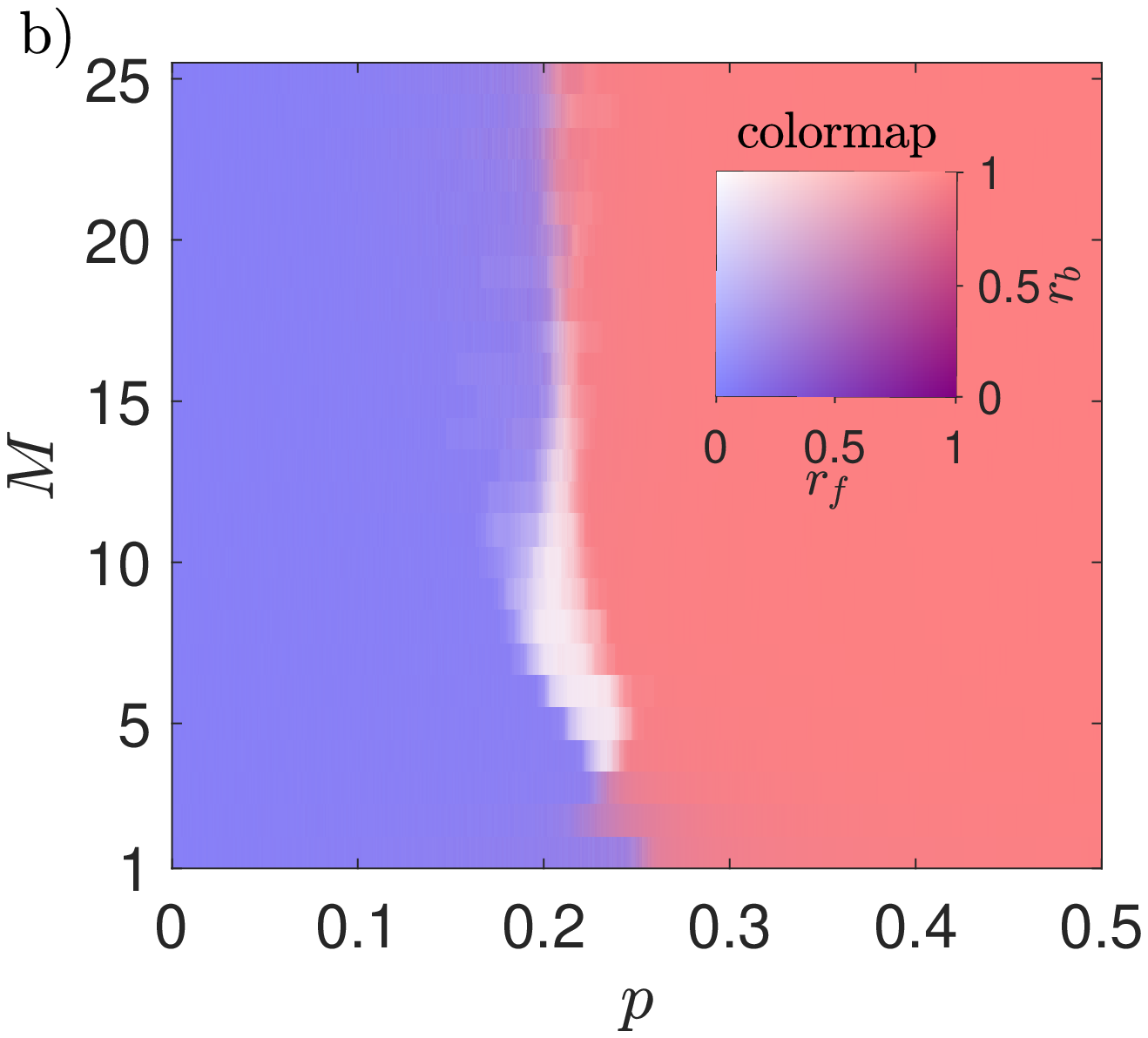}}
\caption{a) Synchronization curves depending on $p$ for $M = 2$ (black) and $M = 10$ (red). Measurements are taken at every single link. Parametrization is the same as in Fig. 1 of the main text, but coupling strength is set to $\lambda = 0.025$. b) Phase-space in the $(p,M)$. Colormap shows the corresponding values of $r$ in the forward and backward directions. Results are averaged over 25 realizations of the process.}
\label{fig:SuppDir}
\end{figure}

We wonder to which extent the explosive synchronization transitions found in our model remain present in the directed scenario. We numerically find that the bomb-like transitions indeed occur at single directed link changes when the rule of Eq. (\ref{eq:directed-rule}) is applied. In Fig. \ref{fig:SuppDir}.a), we plot two examples of the synchronization curves in the forward and backward directions for two values of $M$, and we observe an abrupt synchronization diagram, with its associated hysteresis, that occurs for a sampling parameter $M = 10$, but it is completely absent for a much lower value $M = 2$ (close to random directed percolation). In Fig. \ref{fig:SuppDir}.b) we show the phase-space depending on both the density $p$ and sampling or noise $M$. Interestingly, we see that, for the coupling value $\lambda = 0.025$, the hysteresis window is quite small, and hysteresis behavior is only present for $M \in [5,20]$. Nevertheless, the phenomenology is qualitatively similar to the undirected case (see Fig. 3 of main text for a proper comparison). 

These results confirm that the synchronization bomb can be extended to directed networks, which may represent a more realistic scenario, at least in biological systems as the brain. A more detailed theoretical study of the directed synchronization bomb, including the analyses of the structural properties such as the percolation threshold or frequency-degree correlations, and its extension to other dynamical processes, is left for further work. Furthermore, the already known appearance of the Braess' Paradox in directed networks \cite{arola21} points towards the counter--intuitive possibility of designing \emph{reversed} synchronization bombs, where the transition from incoherence to global synchrony (or vice-versa) is induced by the removal (or addition) of a single directed link.

\bibliography{main.bib}

\end{document}